\documentclass[iop]{emulateapj}

\newcommand{\lya}{Ly$\alpha$~}

\newcommand{\nhi}{N(\mbox{H~{\sc i}})}
\newcommand{\lognhi}{\log~\nhi}

\newcommand{\pcmsq}{cm$^{-2}$}
\newcommand{\cii}{C~{\sc ii}~}

\newcommand{\civ}{C~{\sc iv}~}

\newcommand{\oi}{O~{\sc i}~}

\newcommand{\siii}{Si~{\sc ii}~}

\newcommand{\siiv}{Si~{\sc iv}~}
\newcommand{\hi}{H~{\sc i}~}

\newcommand{\mginsp}{Mg~{\sc i}}
\newcommand{\mgiinsp}{Mg~{\sc ii}}
\newcommand{\feiinsp}{Fe~{\sc ii}}

\newcommand{\aliinsp}{Al~{\sc ii}}
\newcommand{\aliiinsp}{Al~{\sc iii}}
\newcommand{\siiinsp}{Si~{\sc ii}}
\newcommand{\siivnsp}{Si~{\sc iv}}
\newcommand{\civnsp}{C~{\sc iv}}

\newcommand{\oinsp}{O~{\sc i}}

\newcommand{\hinsp}{H~{\sc i}}

\newcommand{\ciinsp}{C~{\sc ii}}

\newcommand{\feii}{Fe~{\sc ii}~}
\newcommand{\alii}{Al~{\sc ii}~}
\newcommand{\aliii}{Al~{\sc iii}~}
\newcommand{\mgi}{Mg~{\sc i}~}
\newcommand{\mgii}{Mg~{\sc ii}~}

\newcommand{\ang}{\AA~}
\newcommand{\angmath}{\textrm{\AA}}

\newcommand{\kms}{km s$^{-1}$}
\newcommand{\mgiiwr}{W_{0}^{\lambda2796}}
\newcommand{\hiwr}{W_{0}^{\lambda1215}}
\newcommand{\feiiwr}{W_{0}^{\lambda2600}}
\newcommand{\civwr}{W_{0}^{\lambda1548}}
\newcommand{\mgiiks}{\omega_{\lambda2796}}

\newcommand{\nzero}{66}
\newcommand{\nmgiiorig}{393}
\newcommand{\nmgiisys}{313}
\newcommand{\nprox}{14}
\newcommand{\nsysorig}{2705}
\newcommand{\zmincomp}{0.01}
\newcommand{\zmaxcomp}{2.44}
\newcommand{\wmincomp}{0.030}
\newcommand{\wmaxcomp}{5.796}
\newcommand{\kmssame}{250}
\newcommand{\kmsqso}{10,000}

\newcommand{\fireexpscale}{0.824}
\newcommand{\nestorexpscale}{0.702}
\newcommand{\nlow}{272}
\newcommand{\zlow}{1.128}
\newcommand{\nhigh}{97}
\newcommand{\zhigh}{3.184}
\newcommand{\nsdss}{1975}
\newcommand{\zsdss}{1.064}

\newcommand{\nlowthree}{217}
\newcommand{\zlowthree}{0.968}
\newcommand{\nmidthree}{100}
\newcommand{\zmidthree}{2.092}
\newcommand{\nhighthree}{52}
\newcommand{\zhighthree}{3.780}

\newcommand{\pdlazmin}{0.613}
\newcommand{\pdlazmax}{4.282}
\newcommand{\pdlazmean}{3.019}

\newcommand{\ndla}{62}
\newcommand{\dlalogrank}{98.2}
\newcommand{\dlagehhyp}{69.8}
\newcommand{\psubdlazmin}{2.076}
\newcommand{\psubdlazmax}{4.244}
\newcommand{\psubzmean}{3.273}
\newcommand{\ndsub}{1029}

\newcommand{\npsub}{77}
\newcommand{\prtmax}{2500}

\newcommand{\nrt}{197}
\newcommand{\rtzmin}{0.116}
\newcommand{\rtzmax}{1.645}
\newcommand{\rtwmin}{0.300}
\newcommand{\rtwmax}{3.264}

\newcommand{\rtnmin}{18.18}
\newcommand{\rtnmax}{21.71}
\newcommand{\rtnerr}{0.087}
\newcommand{\nrtsub}{37}
\newcommand{\rtsubzmin}{0.430}
\newcommand{\rtsubzmax}{1.645}
\newcommand{\rtsubzmean}{0.927}

\newcommand{\rtmgeoprob}{82.1}
\newcommand{\rtfgeoprob}{71.1}
\newcommand{\rttmax}{2500}
\newcommand{\rtnbins}{4}

\newcommand{\hiznhiks}{96.7}
\newcommand{\nnhifire}{27}

\newcommand{\nmeanfiresub}{20.16}
\newcommand{\ndlasfire}{11}
\newcommand{\perdlasfire}{40.7}

\newcommand{\nhihisubzmean}{3.402}

\newcommand{\nhiks}{0.008}
\newcommand{\rtperdlas}{16.7}
\newcommand{\rtperdlaslow}{5.3}
\newcommand{\rtperdlashigh}{7.1}
\newcommand{\perdlaslow}{9.2}
\newcommand{\perdlashigh}{9.8}
\newcommand{\nhisyserrprob}{2.0}

\newcommand{\cdfhimeanlow}{2.82}

\newcommand{\cdfhimeanhigh}{8.86}

\newcommand{\cdfaliiiameanlow}{0.18}

\newcommand{\cdfaliiiameanhigh}{0.04}

\newcommand{\cdfwhimeanlow}{6.89}

\newcommand{\cdfwhimeanhigh}{5.44}

\newcommand{\cdfwcivameanlow}{0.97}

\newcommand{\cdfwcivameanhigh}{0.29}

\newcommand{\dividewrsuboneks}{greater than $99.99$}

\newcommand{\dividewrnweak}{119}
\newcommand{\dividewrnstrong}{75}

\newcommand{\magesnrmin}{3.5}
\newcommand{\magesnrmax}{38.5}
\newcommand{\qmagesnr}{23}
\newcommand{\brthreesnr}{43}
\newcommand{\brfoursnr}{32}
\newcommand{\hiressnr}{40}
\newcommand{\sdsssnrmin}{3.2}
\newcommand{\sdsssnrmax}{30.1}
\newcommand{\sdsssampsnr}{14}

\newcommand{\nsdssorig}{17,296}
\newcommand{\sdssorigzmin}{0.366}
\newcommand{\sdssorigzmax}{2.223}
\newcommand{\sdssorigwmin}{0.19}
\newcommand{\sdssorigwmax}{7.98}

\newcommand{\firesnrone}{20}
\newcommand{\firesnrtwo}{26}
\newcommand{\firesnrthree}{47}
\newcommand{\firesnrfour}{12}

\newcommand{\znhim}{0.359}
\newcommand{\znhimerr}{0.081}
\newcommand{\znhib}{18.966}
\newcommand{\znhiberr}{0.153}
\newcommand{\zratm}{2.783}
\newcommand{\zratmerr}{0.706}
\newcommand{\zratb}{1.980}
\newcommand{\zratberr}{1.434}

\newcommand{\chclassnall}{21}

\newcommand{\fireclassnall}{94}
\newcommand{\classper}{84.2}
\newcommand{\ndlamid}{6}
\newcommand{\ndoublemid}{3}

\newcommand{\classnhuge}{8}

\newcommand{\classhugefe}{2.26}

\newcommand{\classhugeks}{128.3}

\newcommand{\classnbigcc}{11}
\newcommand{\hugecut}{2.75}

\begin{document}

\newcommand\advplotone[4]{\plotone{#2}}
\newcommand\advinputin[2]{} 
\newcommand\advinputend[1]{\input{#1}} 
\newcommand\paperone{Paper I}
\newcommand\paperonesp{Paper I ~}
\newcommand\firstpaper{Matejek \& Simcoe (2012; hereafter, \paperone)}

\title{\mgii Absorption at $2<z<6$ with Magellan~/~FIRE. \\ II: A Longitudinal Study of \hinsp, Metals, and Ionization in
  Galactic Haloes \altaffilmark{1}}
\shorttitle{\hinsp, Metals \& Ionization in Galactic Haloes}
\subjectheadings{Galaxies: evolution---Galaxies: halos---Galaxies:
  high-redshift---Infrared: general---intergalactic medium---quasars:
  absorption lines} \author{Michael S. Matejek\altaffilmark{2}, Robert
  A. Simcoe\altaffilmark{2}, Kathy L. Cooksey\altaffilmark{2,3} and Eduardo N. Seyffert\altaffilmark{2}} \altaffiltext{1}{This paper includes data
  gathered with the 6.5 meter Magellan Telescopes located at Las
  Campanas Observatory, Chile}\altaffiltext{2}{MIT-Kavli Institute for
  Astrophysics and Space Research, Massachusetts Institute of
  Technology, 77 Massachusetts Ave., Cambridge, MA 02139, USA}\altaffiltext{3}{NSF Astronomy \& Astrophysics Postdoctoral Fellow}

\begin{abstract}
We present a detailed study of \hi and metals for 110 \mgii absorption
systems discovered at $1.98\le z\le5.33$ in the infrared spectra of
high redshift QSOs.  Using new measurements of rest-frame UV lines
from optical spectra of the same targets, we compare the high redshift
sample with carefully constructed low redshift control samples from
the literature to study evolutionary trends from $z=0\rightarrow5.33$
($>12$ Gyr).  We observe a significant strengthening in the
characteristic $\nhi$ for fixed \mgii equivalent width as one moves
toward higher redshift.  Indeed at our sample's mean
$\bar{z}$=\nhihisubzmean, all \mgii systems are either damped
Ly$\alpha$ absorbers or sub-DLAs, with $\perdlasfire$\% of systems
exceeding the DLA threshold (compared to $\rtperdlas$\% at
$\bar{z}$=\rtsubzmean).  We set lower limits on the metallicity of the
\mgii systems where we can measure \hinsp; these results are
consistent with the full DLA population.  The classical \mgii systems
($\mgiiwr=0.3-1.0$\AA), which preferentially associate with sub-DLAs,
are quite metal rich at $\sim 0.1$ Solar.  We applied quantitative
classification metrics to our absorbers to compare with low redshift
populations, finding that weak systems are similar to classic \mgii
absorbers at low redshift.  The strong systems either have very large
\mgii and \feii velocity spreads implying non-virialized dynamics, or
are more quiescent DLAs.  There is tentative evidence that the
kinetically complex systems evolve in similar fashion to the global
star formation rate.  We speculate that if weaker \mgii systems
represent accreting gas as suggested by recent studies of
galaxy-absorber inclinations, then their high metal abundance suggests
re-accretion of recently ejected material rather than first-time
infall from the metal-poor IGM, even at early times.
\end{abstract}

\section{Introduction}
\label{sec:ions_intro}

For decades, \mgii quasar absorption lines have been used to probe the
gas distribution in $z<2.3$ galactic haloes in a largely dust
extinction and luminosity-independent manner \citep[e.g.,
][]{weymann_mgii,lanzetta_mgii,tytler_mgii,sargentmgii,steidelandsargent,nestor2005,prochtersdss2006,lund2009}.
Despite this rich literature, the spatial structure and dynamical
history of the gas giving rise to \mgii absorption are not fully
understood.  Several important clues have surfaced through the
aforementioned studies, and (broadly speaking) they point to two
plausible mechanisms.  The first possibility is that \mgii traces cool
clumps embedded in hot galactic outflows \citep[e.g.,
][]{zibetti2007colors,bouche2007ha,ubiquitous2009,gauthier,lund2009,rubin2010,noterdaeme2010,menardo2}.
The second postulates that \mgii absorbing structures are a
manifestation of gravitational and gas accretion processes, perhaps
even through recycled and metal-enriched winds \citep[e.g.,
][]{chen2010lowz,chen2010sfr,lovegrove2011,kacprzak2011incl}.

The outflow hypothesis is supported by low redshift studies showing a
connection between \mgii absorption and star formation.  For example,
\citet{zibetti2007colors} demonstrated that strong absorber $\mgiiwr$
correlates with blue host galaxy color, using a sample of 2800 strong
\mgii systems ($\mgiiwr>0.8\textrm{\AA}$) at low redshifts
($0.37<z<1.0$), a result later corroborated by \citet{lund2009}.

More directly, \citet{ubiquitous2009} observe blueshifted (and hence
outflowing) foreground \mgii absorption in the stacked spectra of star
forming galaxies.  Follow up work by \citet{rubin2010} verifies this
trend and establishes a correlation between \mgii rest-frame
equivalent width and star formation rate (SFR).
\citet{nestor2010strong} studied two ultra-strong \mgii absorbers
($\mgiiwr$ = 3.63 and 5.6 \AA) in detail, finding that they were
associated with galaxies of unusually high specific star formation
rate at their respective masses and redshifts.

Further evidence of a \mgiinsp-wind connection may be found from 
studying statistical clustering of \mgii systems relative to nearby galaxies.
\citet{gauthier}, following up on the work of
\citet{bouche2006anticorrelation} and \citet{lund2009}, find a
$1\sigma$ anti-correlation between \mgii rest-frame equivalent width
and galaxy halo mass by cross-correlating luminous red galaxies with
$\mgiiwr\gtrsim1.0\textrm{\AA}$ \mgii absorbing systems from SDSS-DR5
at $z\sim 0.5$.  Although the anti-correlation is weak, in conjunction with studies showing a strong correlation between $\mgiiwr$ and velocity spread \citep{ellisonEW,matejek2012} it suggests that
the individual \mgii systems are not virialized.

While these and other studies \citep[e.g.,
][]{bouche2007ha,noterdaeme2010,menardo2} have advocated outflows as a
mechanism for creating \mgii absorption, there is also evidence
suggesting that many \mgii systems do not originate in winds.  For
example, \citet{chen2010lowz} find little evidence for correlation
between absorber strength and galaxy colors using a galaxy-selected
sample of \mgii systems, in direct contrast to
\citet{zibetti2007colors}.  Similar galaxy-selected samples of weaker
absorbers confirm this result \citep{lovegrove2011,kacprzak2011incl}.
Moreover, \citet{chen2010sfr} demonstrate with a sample of 47 weaker
(mostly $\mgiiwr < 1$ \AA) systems at $z<0.5$ that the extent of the
\mgii halo correlates only weakly with specific star formation rate
and increases with galaxy stellar mass.  The authors suggest that this
may be evidence that \mgii absorbers reside in infalling clouds that
later fuel star formation.

Recently, \citet{bordoloi} and \citet{kacprzak2011incl} have explored
the connection between absorber strength and galaxy-absorber projected
inclination, finding evidence for both co-planar and bipolar
distributions of absorbing gas.  While the outflow hypothesis
naturally predicts winds escaping perpendicular to galactic disks as
found by \citet{bordoloi}, the analysis of \citet{kacprzak2011incl} indicates that co-planar gas
exists around some systems, as might be found in accreting streams and
filaments.

Collectively, these studies seem to suggest that \mgii absorbers fall
into at least two categories, as outlined in \citet{kacprzak2011schech}.  Loosely
speaking, weaker absorbers $\mgiiwr\lesssim1\angmath$ are more likely
to possess disk-like kinematics and trace infalling or recycled
material.  The stronger absorbers $\mgiiwr\gtrsim1\angmath$ have
non-gravitational kinematics and are more likely to trace winds.
However all of these results were derived from relatively low redshift
($z<2$) systems that postdate the star formation peak of the universe
at $z\sim2.5-3$.  Since star formation plays an important role in this
discussion, the evolution of \mgii absorbers through the rise and fall of the
SFR history provides a diagnostic tool for evaluating the two-sample
paradigm.  But \mgii absorption at these higher redshifts falls into
the near infrared, where atmospheric OH emission and telluric
absorption make large systematic surveys much more difficult.

In \firstpaper, we presented the first
statistically characterized sample of \mgii absorption lines at
$z>2.5$, taken from the spectra of 46 QSO sightlines observed with
Magellan/FIRE.  We located 110 intervening \mgii systems (plus one
proximate system) ranging in rest equivalent width from
$\mgiiwr=0.08\angmath$ to $\mgiiwr=5.58\angmath$ and in redshift from
$z=1.98$ to $z=5.33$.  The weaker $\mgiiwr < 1 \angmath$ systems'
linear density $dN/dX$ is statistically consistent with no evolution
from $z=0.4$ to $z=5.5$ (a span of over 8 Gyr).  In contrast, the
stronger $\mgiiwr > 1 \angmath$ systems' linear density increases
three-fold until $z\sim3$ before declining again towards higher
redshifts.  The evolutionary behavior of these strong systems suggests
that there may indeed be a connection between star formation and the
strong end of the \mgii population.

The present study follows up the initial survey of \paperonesp by studying
the full properties of each individual $z>2$ \mgii system in detail.
Combination of our data with multiple low redshift samples yields a
longitudinal view of \hi and metals in \mgiinsp-selected absorbers over a
wide baseline in redshift.  For the $z>2$ sample, we also benefit from
the shifting of vacuum ultraviolet lines including \hinsp, \civnsp, and
other baseline metal transitions into optical wavelengths.  This
allows us to leverage a large assortment of ground based measurements
to study the systems' chemistry and ionization.

Our goals in investigating the internal properties of individual \mgii
systems over a wide time baseline are: (1) to determine whether the
lack of evolution in $dN/dX$ for weak systems (found in \paperone)
reflects a truly non-evolving population or rather masks internal
evolution that is manifested in other observables; (2) to determine
whether the dichotomy between outflowing and infalling \mgii is
revealed in properties other than $\mgiiwr$, such as chemical
composition or \hi column density; and (3) to develop a taxonomy for
high redshift systems and determine how these relate to low-redshift
classes of \mgii systems and in what proportions.

Section \ref{sec:ions_data} describes our sample data.  In Section
\ref{sec:ions_analysis}, we describe our data analysis techniques,
detailing our calculations of metal rest equivalent widths, column
densities, metallicities, and velocity spreads.  In Section
\ref{sec:ions_results}, we present our main science results, including all
measured values, correlations, and Kaplan-Meier/K-S test results.

In Section \ref{sec:ions_discuss}, we discuss the implications of these
results to the broader question of \mgii absorption.  In particular, in Section \ref{sec:ions_classify} we apply a quantitatively derived taxonomy based upon that in \citet{churchill_zoo} and study the evolution of various classes.  In Section \ref{sec:ions_dlas}, we compare the \mgiinsp-selected DLA population to the full population.  In Section \ref{sec:ions_chemevol} we discuss possible interpretations resulting from our chemical composition study.

Throughout this paper we use a $\Lambda$CDM cosmology with
$\Omega_m=0.3$, $\Omega_\Lambda=0.7$, and $H_0=70$ km s$^{-1}$
Mpc$^{-1}$.

\section{Data Sample}
\label{sec:ions_data}

Our overall analysis contains a large number of heterogeneous
subsamples both observed by our group and collected from the
literature, yielding a total sample of over 17,500 absorbers ranging from
$0<z<5.3$.  In Sections 2.1 and 2.2 we describe the infrared and
optical observations of the $z>2$ systems obtained by our group for
the primary survey.  Section 2.3 and associated subsections describe
the numerous samples collected from the literature that serve as our low
redshift control.

\subsection{The FIRE \mgii sample}

\paperonesp provides the full details of the acquisition and reduction of
this data.  Briefly, we observed 46 QSO sightlines with Magellan/FIRE
\citep{fireDesign2008,fireDesign2010}, between 2010 June and 2011
April.  FIRE is a single object, prism cross-dispersed infrared
spectrometer with a FWHM spectral resolution of $\sim 50$ \kms.  The
survey quasars have emission redshifts between 3.55 and 6.28, and were
predominantly chosen from the SDSS DR7 quasar catalog
\citep{schneider}, although some bright, well-known objects not in the
SDSS catalog were also included.

We reduced the data using a custom-developed IDL pipeline named
FIREHOSE that evolved from the optical echelle reduction software
package MASE \citep{mase}.  We corrected for telluric absorption
features by obtaining spectra of A0V stars at comparable observing
times, air masses and sky positions as our observed QSOs and employing
the xtellcor software package \citep{cushing2004spextool}.  The final
spectra ranged in median signal-to-noise ratio per pixel from 4.0 to
47.2, with a median value of 12.9.

Using automated techniques with interactive verification, we
identified 110 isolated \mgii absorbers ranging in rest equivalent
width from $\mgiiwr=0.08\textrm{\AA}$ to $\mgiiwr=5.58\textrm{\AA}$
and in redshift from $z=1.98$ to $z=5.33$.  We carefully characterized
both the sample's completeness as a function of $\mgiiwr$ and also its
expected false positive rate, adjusting our linear density $dN/dX$
calculations accordingly.  Using the supporting optical data compiled
for the present paper, we have identified two systems from \paperonesp
likely to fall among these false positives, discussed below.  As
expected, they are among the weakest systems in the original sample
($\mgiiwr<0.20$ \ang).  These absorbers are left out of statistical
analysis for both papers because of incompleteness at $<0.3$ \AA; they
are also identified accordingly in all tables presented here.

\subsection{Supporting Optical Spectra}

Rest-frame UV transitions such as \lya and numerous carbon,
silicon, and aluminum transitions are redshifted into the optical
window for $z>1.7$ absorption systems, making these measurements
easily accessible from the ground.  We obtained new or archival
optical spectra for 39 of the 46 QSO sightlines in our survey using
data from four different instruments.  Table \ref{tab:optical}
provides a full description of the optical data, including exposure
times and wavelength coverages.  We limited our metal line search to
regions redward of each QSO's \lya emission peak, and only
searched for \lya absorption redward of the Lyman break for the
highest redshift Lyman limit absorber.  These requirements set the
minimum search wavelength in all our spectra even when the data
extended further to the blue.

\subsubsection{Magellan/MagE - 15 spectra}
We obtained optical spectra of 15 objects with Magellan/MagE, a
single-object echellette \citep{Marshall}, between 2009 March and 2011
January.  We used a 0.7" slit and observed mostly at low airmass in
0.6" to 0.8" seeing.  The spectra were reduced using the MASE pipeline
\citep{mase}.  The 1D spectra range in signal-to-noise per pixel
redward of the \lya Forest from \magesnrmin~ to \magesnrmax, have a 
resolution of $\Delta v = 62.1$ \kms, and span $\lambda \sim 3050$ \ang
to 10280 \AA.

\begin{figure*}
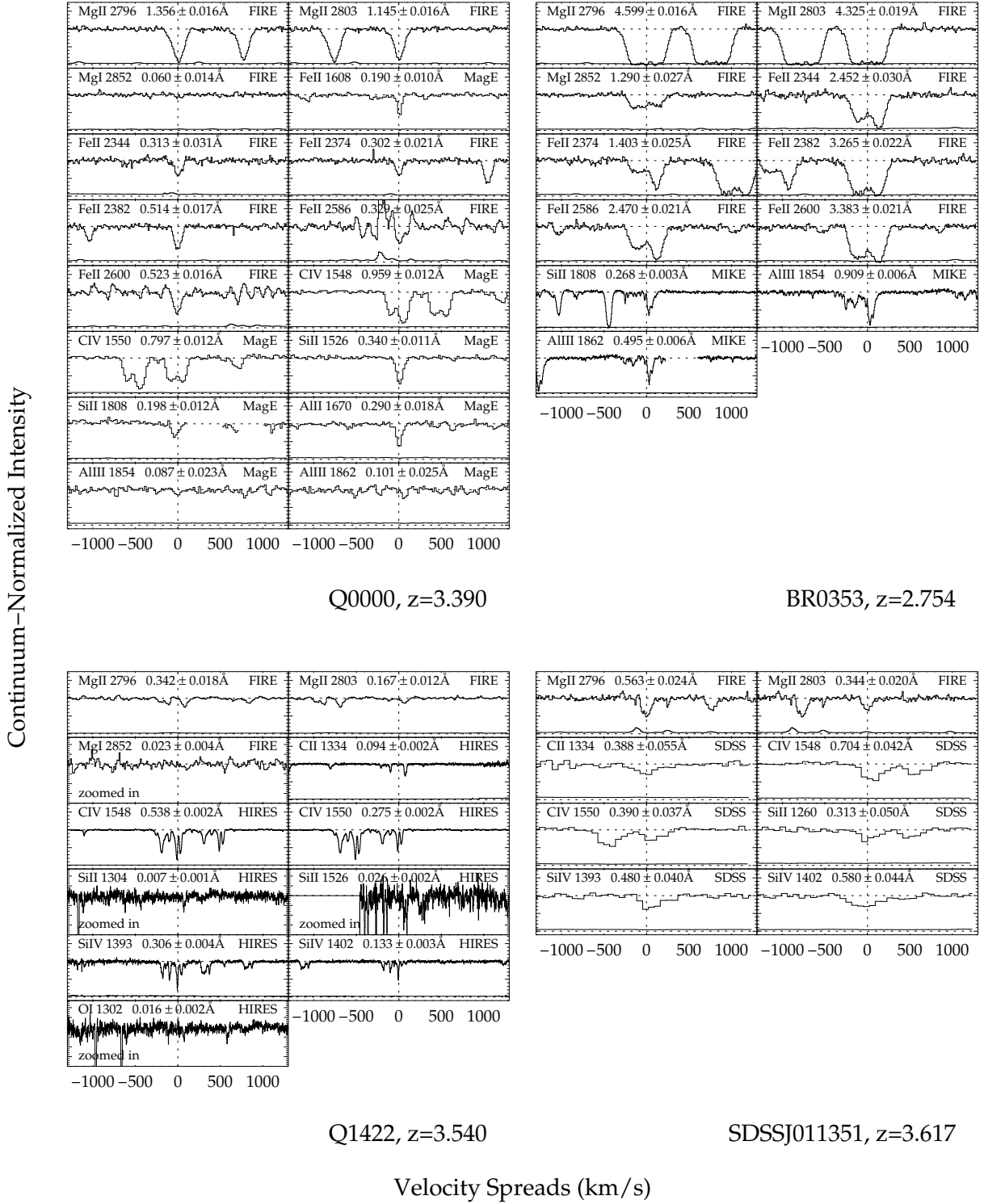

\epsscale{1.15}
\advplotone{Ions}{ion_vel_plots}{7in}{6.5in}
\caption{Samples of all absorption lines detected at a 3$\sigma$ level for 4 of the
  \mgiinsp-selected systems, with the normalized errors overplotted.  The
  horizontal dotted lines lie at zero flux and at the normalized
  continuum, and the vertical dotted line coincides with the zero
  velocity point of the \mgii 2796 transition.  The median
  signal-to-noise ratios per pixel for the FIRE spectra are
  $\sim\firesnrone$, \firesnrtwo, \firesnrthree, and \firesnrfour~
  from left to right, top to bottom.  For the optical counterparts,
  the corresponding signal-to-noise ratios redward of the \lya
  emissions from the QSOs are $\sim\qmagesnr$, \brthreesnr, \hiressnr,
  and \sdsssampsnr.  The approximate resolutions of the instruments
  are 50 \kms~ for FIRE, 62.1 \kms~ for MagE, 14 \kms~ for MIKE, 6.6
  \kms~ for HIRES, and $\sim150$ \kms~ for SDSS.  The \mgi 2852, \siii
  1304, \siii 1526, and \oi 1302 plots for the $z=3.540$ Q1422 system
  are zoomed-in so that the bottom edge lies at 80\% of the continuum
  level.  Nearby lines from other systems were masked when deemed
  distracting.}
\label{fig:ion_vel_plots}
\end{figure*}

Representative regions of the MagE spectrum for Q0000-26 are shown at
the upper left of Figure \ref{fig:ion_vel_plots}, which displays all
metal lines detected at a 3$\sigma$ level in the FIRE and MagE spectra at $z=3.390$.  The MagE
spectrum for this object has a median signal-to-noise ratio of
$\sim\qmagesnr$ per pixel redward of the \lya forest.

In the same MagE spectrum, we did not find a $0.162$ \ang \mgii
absorber at $z=2.184$, where the FIRE and the MagE spectra overlap, as
reported in \paperone.  Since the MagE spectrum has a higher signal-to-noise
ratio in this region, we now regard this as a false positive.

\subsubsection{MIKE - 2 spectra}
For two sample quasars (BR0353-3820 and BR0418-5723) we had
high-resolution optical spectra available from previous studies of
\civ for other programs \citep{z4civ}.  These were taken with the
Magellan Inamori Kyocera Echelle \citep[MIKE,][]{mike}, between 2004
and 2006.  The MIKE spectra have a resolution of 14 \kms, span the
wavelength range $\sim 4900$ \ang to 9400 \AA, and have median
signal-to-noise ratios per pixel of $\sim$ \brthreesnr~ for BR0353-3820
and $\sim$ \brfoursnr~ for BR0418-5723.

The upper right set of plots in Figure \ref{fig:ion_vel_plots}
displays all metal lines detected at a 3$\sigma$ level for the $z=2.754$ absorbing system
along BR0353-3820, and includes representative samples of the
BR0353-3820 MIKE spectrum.

\subsubsection{HIRES - 1 spectrum}
Q1422+2309 is a well known gravitationally lensed
quasar. \footnote[1]{The lensing galaxy falls at $z=0.338$
  \citep{q1422lens}, well below the redshift search range for our
  study.  The inclusion of this quasar therefore does not bias our
  results in this work or in \paperone.} We obtained a high ($\sim$
\hiressnr) SNR HIRES \citep{hires} optical spectrum of it from
A. Songaila's spectral archive at the University of
Hawaii\footnote[2]{http://www.ifa.hawaii.edu/users/acowie/spectra/spectra\_hires.html}.
The spectrum, originally published in \citet{ellison_q1422}, has a
resolution of 6.6 \kms~ and covers the wavelength range $\sim 4000$
\ang to 7300 \AA.  All metal lines detected at a 3$\sigma$ level for
the $z=3.540$ system along Q1422+2309 found in both its FIRE and HIRES
spectra are shown in the bottom left set of plots in Figure
\ref{fig:ion_vel_plots}.  The high SNR and resolution of this HIRES
spectrum allow us to detect \siii 1304 absorption of only 7 m\AA.  It
also revealed a greater velocity width for this system than previously
reported in \paperone.  The updated $\mgiiwr$ and $W_0^{\lambda2803}$
values used in this study are provided in Table \ref{tab:mglist}.  (These adjustments do not effect the $dN/dX$ calculations from \paperone~ because this system was missed by our automated finder, and therefore was left out of those calculations to avoid overcompensating for incompleteness.)

\subsubsection{SDSS - 21 spectra}
We downloaded optical counterparts for 21 of the remaining 28 QSO
sightlines from the DR7 spectral
archives\footnote[3]{http://das.sdss.org/www/html/} of the Sloan Digital Sky Survey 
\citep[SDSS,][]{sdssspectra}.  These spectra have a resolution of $\sim 150$
\kms, a wavelength range of $\sim 3800$ to 9250 \AA, and signal-to-noise 
ratios per pixel that range from \sdsssnrmin~ to \sdsssnrmax.
The lower right set of plots in Figure \ref{fig:ion_vel_plots}
includes representative samples of the SDSS spectrum of SDSSJ011351
(signal-to-noise $\sim$ \sdsssampsnr), displaying all metal lines 
detected at a 3$\sigma$ level for the $z=3.617$ absorption system found in both its SDSS and
FIRE spectra.  In general the SDSS spectra are sensitive only to the
stronger metal-line systems, but they are very useful for measuring
\hi column densities.

\subsection{Comparison Samples}
\label{sec:ions_compare}

Since our primary goal is to study the redshift evolution of
\mgiinsp-selected systems, we must also establish a local control sample.
For this purpose we consider four comparison sets: a compilation of
previously published low redshift metal absorption lines, new
measurements of metal lines from a \mgiinsp-selected sample of the SDSS
DR7, a low redshift \emph{HST} sample including $\nhi$ measurements for
\mgiinsp-selected systems, and a damped \lya (DLA) selected
metallicity sample from the literature.

\subsubsection{Low Redshift Literature Compilation}
\label{sec:ions_lowz_compare}

We conducted an extensive compilation of low redshift ($z<2.5$) metal
absorption lines previously reported in the literature to complement
our high redshift survey
\citep{Young1979,sargent1979,ysb1982a,ysb1982b,wright1982,sargent82a,sargent82b,RS1983,Foltz1986,lanzetta_mgii,tytler_mgii,sargentmgii,sargent89,steidel90,petitjean1990,barthel90,steidelandsargent,hst_mgii1,aldcroft94,petitjean94,hst_mgii2,sl1996,hst_mgii3,churchill_weakmgii,churchill_2000a,churchill_zoo}.
For this exercise, we included only blind searches of QSOs not
selected with any prior knowledge about absorption properties.
Because we are studying multiple metal and \hi transitions for each
\mgii system, we favored surveys that reported all detected
transitions, and not simply \mgii (or \mgii and \feiinsp).  In many
cases the same object was observed in multiple surveys covering
different wavelengths and transitions.  To avoid duplication in such
instances, we considered absorption systems whose redshifts matched
within \kmssame~ \kms~ to be the same.

For consistency, we converted all absorption features detected at less than a 
5$\sigma$ significance to upper limits and adjusted all reported 3$\sigma$ 
and 4$\sigma$ upper limits to a 5$\sigma$ level.  Unfortunately, most of 
these surveys do not list upper limits at
the expected locations of undetected transitions.  This omission
becomes important when we attempt to build distribution functions of
$W_r$ for each transition using survival analysis.  To capture this
information in a very conservative way, we estimated upper limits for
all unreported transitions that could have been detected in each
spectrum given its wavelength bounds.  This process is necessarily
crude because we did not have access to the original data, but
ignoring the effect would bias our $W_r$ distributions to the high
side.  For each non-detection, we simply assigned an upper limit equal
to 5 times the largest error listed for an identified absorption line
in that QSO's spectrum.  All lines flagged as blends were also treated
as upper limits.  These blended upper limits, however, violate the principle of random censorship because their values are dependent upon the actual line strengths, and were therefore omitted from survival statistics \citep{unisurvival}.

In all, we located \nsysorig~ unique absorption systems across the
surveys listed above. Of these, \nmgiiorig~ had \mgii2796 absorption
lines detected at more than a 5$\sigma$ significance.  Within this
\mgii subset, we threw out \nzero~ $z\sim0$ systems and \nprox~
proximate systems (which we defined as residing within \kmsqso~ \kms~
of the QSO).  This left us with \nmgiisys ~isolated, \mgiinsp-selected
systems in our compilation set. These systems range in redshift from
\zmincomp ~to \zmaxcomp ~and in rest equivalent width from \wmincomp\AA 
to \wmaxcomp\AA.  Table \ref{tab:lowzsample} contains
measurements for a selection of these transitions on a system by
system basis.

\subsubsection{SDSS DR7 \mgii sample}
\label{sec:ions_dr7_compare}

To augment our low redshift data from the literature, we also searched
for multiple metal line transitions coincident with \mgii systems
identified in spectra from the SDSS DR7 (Seyffert, et al., in prep).  Many previous \mgii absorption studies \citep{nestor2005,prochtersdss2006,quiderSDSS} have worked with SDSS spectra, making this a nice comparison set.  The SDSS DR7 
parent sample includes over 65,000 \mgii systems discovered by an
automated continuum fitting and search algorithm and then
interactively inspected for final approval.  The details of this
process may be found in Seyffert, et al. (in prep) and \citet{cooksey_civ}.

We only considered the subset of systems from this full set which had the
highest possible user-rating on a 4-point scale and $\mgiiwr$ observed at a 
5$\sigma$ significance.  Using these
redshifts, we re-fit a selection of metal transitions in an automated
fashion, and recorded upper limits where no absorption was detected.
The final subset included \nsdssorig~ \mgii absorption systems with $\sdssorigzmin \leq
z \leq \sdssorigzmax$ and $\sdssorigwmin \angmath \leq \mgiiwr \leq
\sdssorigwmax \angmath$.

Although this sample contains many more systems than the low redshift
compilation discussed last section, the SDSS spectra typically have
lower signal-to-noise ratios and are largely incomplete for rest
equivalent widths $\lesssim 1.0\angmath$.  In addition, the automated
determination of metal line rest equivalent widths leaves the sample
vulnerable to continuum errors, blended lines, and other effects
typically spotted and adjusted for during interactive inspection.  For
these reasons we use the DR7 sample as a supplement to the low
redshift comparison set rather than its replacement, despite its
large number of systems.

\subsubsection{\emph{HST} \mgiinsp-selected $\nhi$ Sample}
\label{sec:ions_hst_compare}

One of our chief aims is to characterize the \hi properties of \mgii
systems, but at low redshift only a small fraction of all known \mgii
absorbers have \hi measurements since the \lya transition may only be
observed from space.  \citet{rt2006} present the largest such sample,
with $\nhi$ measurements of \nrt~ systems taken with the \emph{Hubble Space
Telescope} (\emph{HST}), as part of a search for low redshift DLAs.  Their
survey pre-selects based on \mgii strength of known absorbers
identified from a broad literature search (the full list is given in
their Table 1).  Special preference was given to systems with large
\feii 2600 equivalent width to maximize the yield of DLAs.  These
\mgiinsp-selected systems range from $z=$\rtzmin~ to \rtzmax~ and
$\mgiiwr=$\rtwmin~ to \rtwmax \AA.  The $\lognhi$ measurements range
from \rtnmin~ to \rtnmax~ \pcmsq~ with a mean error across the sample
of \rtnerr~ \pcmsq.  

Because this sample was selected specifically to maximize the
probability of uncovering DLAs, it is not a statistically
representative collection of random \mgii systems.  However, it is the
largest \hinsp+\mgii compilation known.  So, we adopt it below and
correct the distribution in postprocessing to make it statistically
equivalent to a randomly drawn \mgii population (Section
\ref{sec:ions_rtsub}).

\subsubsection{Metallicity sample}
\label{sec:ions_dla_compare}

Another topic of interest is whether the metallicities of
\mgiinsp-selected systems differ from the general population at high
redshift. Since most \mgiinsp-selected systems are DLAs or sub-DLAs, we
use \citet{prochaska_dla} as a comparison set.  The authors provide
abundance measurements on 86 DLAs found along 42 QSO sightlines taken
with HIRES/Keck and 65 QSO sightlines taken with the $R=13,000$
echellete on the Echellete Spectrograph and Imager \citep[ESI, ][]{esi}.
The DLA absorption redshifts range from \pdlazmin~ to \pdlazmax~ with
a mean of \pdlazmean.  These systems were not selected for \mgiinsp, but
constitute a high redshift abundance reference.

\section{Analysis}
\label{sec:ions_analysis}

\subsection{\mgii Line Identification}
\label{sec:ions_find}

\paperonesp contains details of the \mgii line identification algorithm
applied to our FIRE data.  Briefly, we used an automated continuum
fitting algorithm, and then ran a matched filter search using a double
Gaussian separated by the \mgii doublet spacing as a kernel.  To
mitigate the high false positive rate caused by intermittent telluric
absorption features and missubtracted emission lines, we subjected
each \mgii candidate to a set of consistency checks (e.g., $\mgiiwr\ge
W_0^{\lambda2803}$ within errors).  Finally, the surviving candidates
underwent a visual inspection before ultimate acceptance.  We fit rest
equivalent widths to each doublet using boxcar summation between
user-defined limits.  

\subsection{Measurements}
\label{sec:ions_ion_measurements}

\subsubsection{Calculating \hi Column Densities}
\label{sec:ions_nhi}

For our high-redshift \mgii systems with \hi coverage, we manually fit
\hi column densities with Voigt profiles using the x\_fitdla routine
in the XIDL
library.\footnote[1]{http://www.ucolick.org/$\sim$xavier/IDL/index.html}
Since all our measured systems turned out to be either DLAs or
sub-DLAs (i.e., above the flat portion of the \hi curve of growth), our
final fits were not sensitive to the $b$ value, and we fixed it at 30
\kms.  This decision was largely a practical consideration, since the
QSOs' high redshifts made it highly likely that unassociated Lyman
limit absorbers would obscure the measurements of Ly$\beta$ or higher
order transitions.

Toward the low end of the $\nhi$ range in our sample, the effect of an
uncertain $b$ becomes more pronounced, so we account for this in the
quoted $\nhi$ errors for these systems.  The resolution of our optical
spectra was typically too low to identify individual subcomponents in
each absorption system, so we fit only one \hi component except for
our HIRES spectrum of Q1422 and two complex systems with wide velocity
spreads.  Even in these cases, the resolution did not allow us to fit unique \hi column densities to the individual components.  Lyman limit absorption from
systems at higher redshift obscured even the \lya transitions
for the majority of our absorption systems; we excluded these from the
\hi sample.

\begin{figure*}
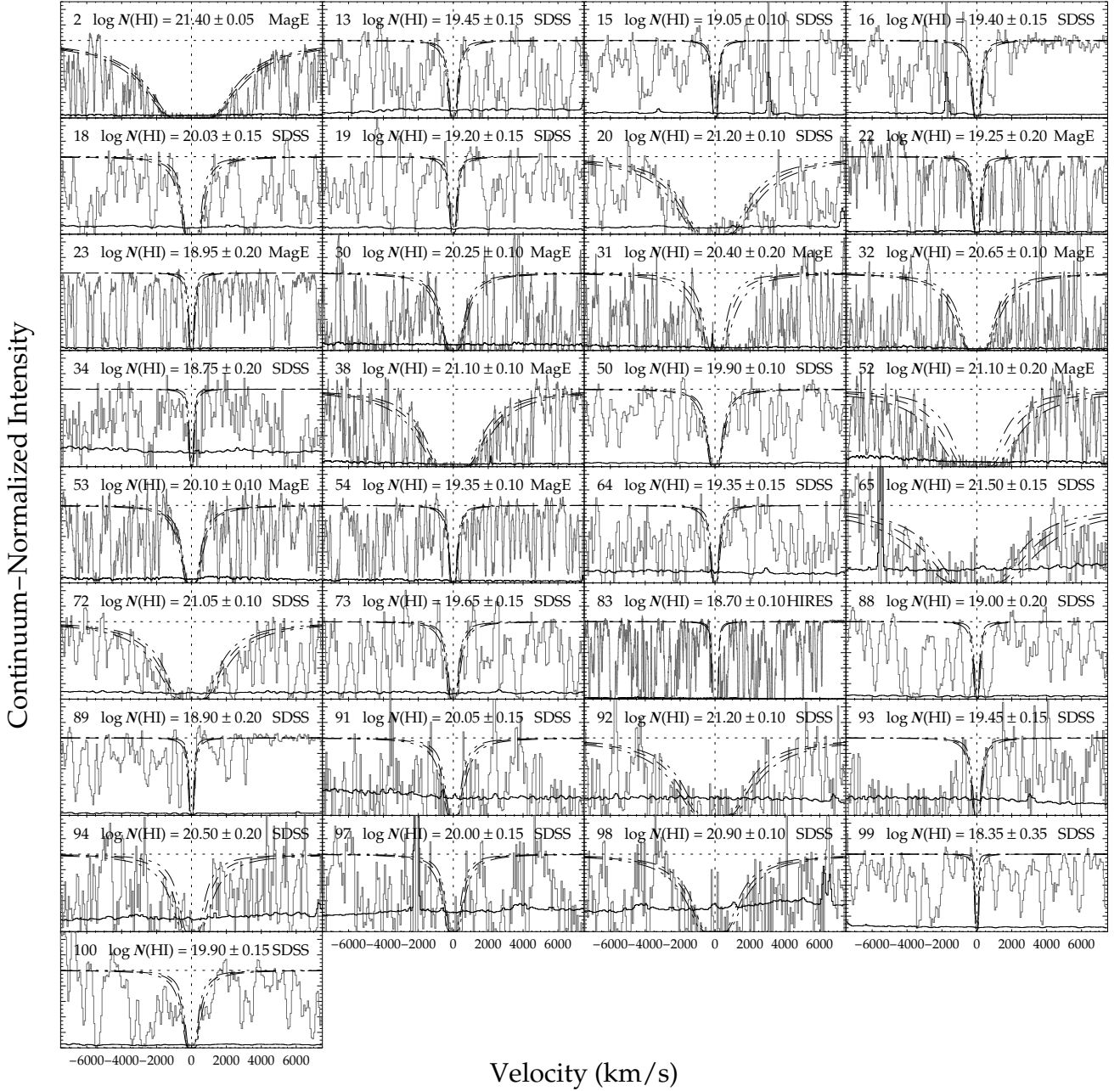

\epsscale{1.15}
\advplotone{Ions}{fire_nhi}{6.5in}{7in}
\caption{\lya absorption profiles for the 33 \mgiinsp-selected high
  redshift systems with $\nhi$ coverage.  The gray line is the
  continuum normalized intensity, and the solid line is the normalized
  error.  The numbers in the upper left are the system index numbers,
  as listed in Table \ref{tab:mglist}.  For reference, the horizontal
  dotted line is normalized continuum, and the vertical dotted line
  rests at the zero velocity point of the \hi profile.  The three
  overplotted lines represent the best fit Voigt profile (dashed) and
  the upper and lower 1$\sigma$ error lines (dashed-dotted),
  calculated using the method described in Section \ref{sec:ions_nhi}.
  The instrument used in each case is given in the upper right.}
\label{fig:nhi}
\end{figure*}

In all, we were able to measure \hi column densities for 33 of the 110
\mgii systems in \paperone.  Plots of these \lya profiles are shown
in Figure \ref{fig:nhi} with their fitted Voigt profiles overplotted.
Table \ref{tab:nhi_table} lists all the measured \hi column densities.

Some surveys from the literature only quote \hi equivalent widths, and
we wished to compare our results with these as well.  We calculated
rest equivalent widths $W_0^{\lambda1215}$ by integrating the area
under the best fit Voigt profiles.  Errors on $W_0^{\lambda1215}$ were
conservatively calculated by employing a boxcar summation of the
normalized error array where the best fit Voigt profile fell below
10\% of the continuum.  These rest equivalent widths are stored in
Table \ref{tab:olist}.

As part of this process, we discovered that the putative \mgii system
at $z=2.825$ toward SDSS0113-0935 (reported in \paperone) exhibited no
\lya in its SDSS spectrum even though the data quality and flux
level should have allowed such a detection.  We therefore consider
this system (with $\mgiiwr=0.194$ \AA) a false positive.

\subsubsection{Metals}
\label{sec:ions_metal}

In addition to the \mgii 2796 and \mgii 2803 rest equivalent widths
calculated in \paperone, we searched redward of the QSOs' \lya
emission peaks in the FIRE and optical spectra for metal transitions at
the locations predicted by the \mgii doublet redshifts.  We employed a
boxcar method to calculate rest equivalent widths for these lines
using interactively-defined limits set to where the flux met the
continuum.  In some cases, no clear absorption lines existed at the
expected locations.  In these cases, we estimated an upper limit by
boxcar summation of pixels.  We made no attempt to disentangle blended
lines, treating such collisions as upper limits.

In this way we fit a large assortment of metal transitions including:
\mgii 2796, \mgii 2803, and \mgi 2852 (Table \ref{tab:mglist}); \feii
1608, \feii 2344, \feii 2374, \feii 2382, \feii 2586, and \feii 2600
(Table \ref{tab:felist}); \siii 1260, \siii 1304, \siiv 1393, \siiv
1402, \siii 1526, and \siii 1808 (Table \ref{tab:silist}); \cii 1334,
\civ 1548, and \civ 1550 (Table \ref{tab:clist}); \alii 1670, \aliii
1854, and \aliii 1862 (Table \ref{tab:clist}); and \oi 1302 (Table
\ref{tab:olist}).

\subsubsection{Kinematic Measurements}
\label{sec:ions_dvs}

We fit velocity spreads $\Delta v$ for all detected metal lines,
except those measured with SDSS spectra (which we omitted because of
their low resolution).  These velocity spreads were calculated by
considering the minimum and maximum wavelengths of the absorption line
(as determined by the user-defined equivalent width limits, where the
absorption line meets the continuum) and correcting for the
instrumental resolution of the spectra.  We conservatively set the
errors on these velocity spreads to be the greater of 10\% and the
pixel width divided by $\sqrt{2}$.

We also measured the ``kinematic spread'' $\omega$ for each
transition, following the analysis of \citet{churchill_zoo}.  This
quantity is defined as the square root of the optical depth-weighted
second moment of the velocity difference from the centroid (their Equation 1).  Table
\ref{tab:ks} contains all measured kinematic and velocity spreads for
the \mgii 2796, \feii 2600, and \civ 1548 transitions for the FIRE
sample.  We substituted measurements
for other transitions when possible if the main transition could not be
measured (e.g., \feii 2586 for \feii 2600).

\subsubsection{Metallicities}
\label{sec:ions_metalmeasurements}

We calculated metallicity values or lower limits for the 33 absorption
systems with measured \hi column densities (Table
\ref{tab:nhi_table}).  First, we estimated column densities for all
detected metal absorption lines using the apparent optical depth (AOD)
 method of \citet{aod}, although the corresponding values represent
lower limits on column densities for saturated lines.  To determine whether an
absorption line was saturated, we estimated the rest equivalent width
at which the curve of growth becomes non-linear for each metal
transition (conservatively setting $b= 5$ \kms).  Since we generally
do not resolve the absorption complexes into their constituent
subcomponents, we used this rest equivalent width threshold as the
barrier between saturation (stronger absorption) and non-saturation
(weaker).  This may overestimate the likelihood of saturation for
lines with significant velocity substructure, but it provides the most
robust possible lower limit.

For each ion (e.g., \feiinsp), we used the average column densities of all non-saturated
transitions (e.g., \feii 1608, 2344, 2374,
2382, 2586, and 2600) and divided by $\nhi$ as determined above.  When
all metal transitions were saturated, we used the highest value of the
lower limit.  Finally we normalized to the solar abundance scale of
\citet{solar_abund}.  We did not apply ionization corrections to the
metallicity estimates.  This approximation is suitable for DLAs, but
may lead to errors at the 0.1-0.3 dex level for lower $\nhi$ systems
in our sample that would be classified as sub-DLAs \citep[ see also
  Section \ref{sec:ions_metals}]{peroux2007}.

\subsection{Normalizing \mgii Samples for Statistical Comparison}
\label{sec:ions_unbiased}
Given the heterogeneous nature of our high redshift and control
samples, we exercised special care to create selected subsamples for
statistical comparisons.  Our goal is to isolate effects that are
intrinsically evolving in the source population, and reduce our
sensitivity to observational and/or selection biases that differ
between samples.

\subsubsection{Generating Low and High Redshift Samples}
\label{sec:ions_subsample}

First, we divided the total set of all \mgiinsp-selected systems
from FIRE and the literature compilation with \mgii 2796 detected at a 5$\sigma$ 
level into a low redshift
and a high redshift sample, separated at $z=2$.  The low redshift
compilation set from the literature is predominantly $z<2$ systems and
the FIRE set is predominantly $z>2$ systems, so these two samples are
roughly exclusive.  We formed a separate $0.36<z<2$ low redshift set
from the SDSS data.

\begin{figure}
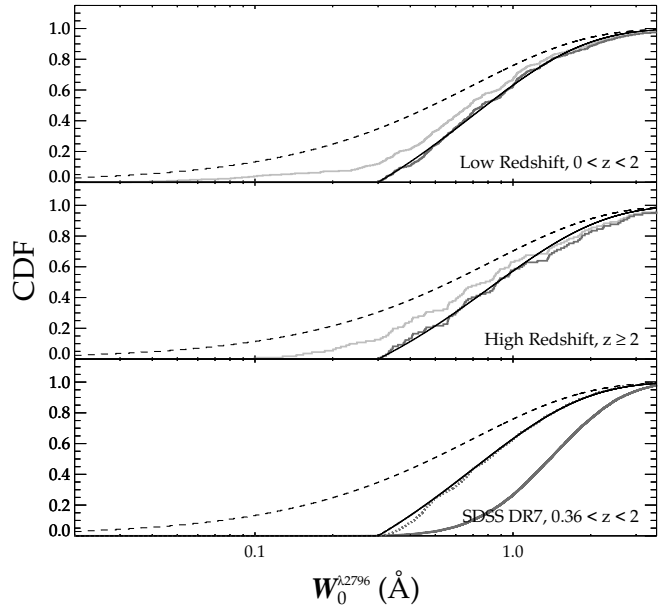

\epsscale{1.20}
\advplotone{Ions}{mgii_cdfs}{5.85in}{5in}
\caption{The CDFs for \mgii rest-frame equivalent width for the low
  redshift ($0<z<2$; upper panel), high redshift ($z\geq2$; middle
  panel), and SDSS DR7 ($0.36<z<2$; lower panel) samples, as described
  in Section \ref{sec:ions_subsample}.  The light gray lines are CDFs
  including all \mgii systems, and the dark gray lines are those under
  the restriction $\mgiiwr\geq0.3\angmath$.  The black lines are
  analytically derived CDFs assuming an exponential frequency
  distribution $dN/dW=\exp(-W/W_*)$ for all (dashed) and $\mgiiwr\geq0.3\angmath$ (solid) systems.  The critical scaling parameters $W_*$ are taken from
  completeness-corrected maximum likelihood estimates in \paperonesp (for
  the high redshift sample) and \citet[][for the low
  redshift and SDSS samples]{nestor2005}.  Making a $\mgiiwr\geq0.3\angmath$ cut
  eliminates the incompleteness evident when considering all systems
  in the low and high redshift samples.  Also plotted as a dotted,
  dark gray line for the SDSS plot is the CDF for the subsample drawn
  to mimic the analytical CDF with $\mgiiwr\geq0.3\angmath$.}
\label{fig:mgii_cdfs}
\end{figure}

Figure \ref{fig:mgii_cdfs} shows the cumulative distribution function
(CDF) of \mgii 2796 rest equivalent width for our low redshift sample
(top panel), high redshift sample (middle panel), and the SDSS
sample (bottom panel).  In each panel, the light gray line includes
all identified \mgii systems, and the dashed black line represents
the analytic CDF for an exponential equivalent width distribution
($dN/dW\propto\exp(-W/W_*)$), with $W_*$ appropriate for each respective
epoch.  For high redshifts we set $W_*$=\fireexpscale, the
completeness-corrected value measured in \paperone.  For both the low
redshift and SDSS DR7 sample, the overplotted exponentials have
$W_*$=\nestorexpscale, which is the completeness-corrected value
from \citet{nestor2005}, measured from the SDSS Early Data Release sample.
The large underabundance of systems with small rest equivalent width
is an indication of incompleteness.

Following past convention from \paperone, we can minimize incompleteness
effects by restricting our analysis to $\mgiiwr\geq0.3\angmath$
systems.  The dark gray lines in Figure \ref{fig:mgii_cdfs}
display the CDFs for this cut, while the solid black curves represent
the same analytically calculated CDFs as before, but with this lower
limit imposed.  The analytic and observed distributions show good
agreement for the high and low redshift subsamples.  The SDSS DR7
sample still suffers from incompleteness issues below $\sim1\angmath$.
Rather than slicing all data to $\mgiiwr\geq1.0\angmath$ (which would
result in very few systems at high redshift), we instead forced the
SDSS set to fit the analytic CDF by drawing an appropriately weighted
subsample from the SDSS parent population.  The resulting CDF of this
subset is overplotted in Figure \ref{fig:mgii_cdfs} as a dotted, dark
gray line.

\begin{figure}
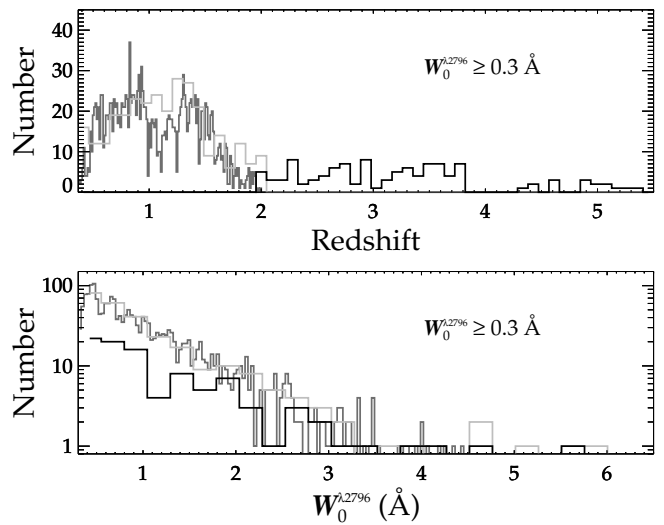

\epsscale{1.20}
\advplotone{Ions}{ion_raws}{6in}{4.85in}
\caption{Histograms in redshift (upper panel) and $\mgiiwr$ (lower
  panel) for the final $0<z<2$ low redshift (light gray line),
  $z\geq2$ high redshift (black line), and $0.36<z<2$ SDSS DR7 (dark
  gray line) samples.  The bin density for the SDSS DR7 sample is eight
  times larger in order to facilitate overplotting.  Section
  \ref{sec:ions_subsample} details the creation of these samples, which are
  restricted to $\mgiiwr\geq0.3\angmath$.}
\label{fig:ion_raws}
\end{figure}

Figure \ref{fig:ion_raws} displays histograms of the redshift (upper
panel) and rest-frame equivalent width (lower panel) for these lower
redshift (light gray line), higher redshift (black line), and SDSS DR7
(dark gray line) samples.  The bins for the SDSS DR7 sample are eight
times smaller in order to facilitate overplotting.  The low $z$, high $z$, 
and SDSS samples contain \nlow, \nhigh, and \nsdss~ systems at mean redshifts 
of \zlow, \zhigh, and \zsdss, respectively.  Table \ref{tab:sample_counts} 
lists the mean redshift, number of
detections and upper limits, and minimum and maximum rest-frame
equivalent widths for a selection of metal transitions in each sample.

Because our low/high redshift cut was motivated chiefly by
observational setup (optical versus IR \mgii measurement), the
designations are arbitrary with respect to any physical evolution.  To
test another prescription, we generated a second redshift
classification with three bins at $0<z<1.5$, $1.5\leq z<3$, and $z\geq3$
using identical methods.  
The final low, mid, and high $z$ samples contained
\nlowthree, \nmidthree, and \nhighthree~ systems at mean redshifts of
\zlowthree, \zmidthree, and \zhighthree, respectively.  Table
\ref{tab:sample_counts3} contains a breakdown of the mean redshift,
number of detections and upper limits, and minimum and maximum rest
frame equivalent widths for various metal transitions for each of
these three redshift groups.

\subsubsection{Generating Weak and Strong $\mgiiwr$ Samples}
\label{sec:ions_wrsubsample}

In \paperone, we presented evidence of differential evolution 
for strong versus weak \mgii absorbers.  For the weak systems
with $0.3\angmath< \mgiiwr<1.0\angmath$, $dN/dX$ is statistically
consistent with no evolution, but for strong $\mgiiwr>1.0\angmath$
absorbers it rises until $z\sim2-3$ and then falls again.  For this
paper we form separate weak and strong $\mgiiwr$ samples to
investigate whether the full chemical compositions of weak and strong
absorbers also differ.

We began with the \nlow~ low redshift and \nhigh~ high redshift
systems from the comparison samples described in Section
\ref{sec:ions_subsample}.  We wanted to avoid our final subsamples
containing disproportionately more low redshift systems, so we only
included a subset of \nhigh~ of these low redshift systems in our
final absorption strength samples.  (This subset has a \dividewrsuboneks\% K-S probability of being drawn from the same $\mgiiwr$ distribution as the full low redshift sample; i.e., it is not an unusual draw).
The final result is a 
weak \mgii absorber sample with $0.3\angmath\leq \mgiiwr
\leq 1.0\angmath$ containing \dividewrnweak~ systems, and a strong
\mgii absorber sample with $\mgiiwr>1.0\angmath$ containing
\dividewrnstrong~ systems.  Table \ref{tab:sample_counts_wr} contains
a breakdown of the mean redshift, number of detections and upper
limits, and minimum and maximum rest-frame equivalent widths for 
these weak and strong \mgii samples.

We additionally constructed a series of four 
subsamples split along both redshift and \mgii absorption strength 
 (high/low redshift, weak/strong equivalent width) with the goal of 
studying the redshift evolution of weak and strong \mgii absorbers.  We created these four 
subsamples by dividing the representative low ($0<z<2$) and high ($z\geq2$) redshift samples of 
Section \ref{sec:ions_subsample} into two groups each, comprising weak ($0.3\angmath\leq \mgiiwr
\leq 1.0\angmath$) and strong ($\mgiiwr>1.0\angmath$) absorbers.  Table 
\ref{tab:sample_counts_wrz} contains the analogous sample count 
information for this set.

\subsubsection{Normalizing Biases in the $\nhi$ Sample}
\label{sec:ions_rtsub}

We use the \emph{HST} $\nhi$ measurements of \citet{rt2006} as a low redshift
\hi control sample, but these authors' primary goal was to locate
DLAs, and \mgii measurements mostly served as a means to this end.
\citet{rt2000} previously showed that 11 of 12 DLAs in their sample
had $\mgiiwr>0.6\angmath$, and more than half of all absorbers with
$\mgiiwr>0.5\angmath$ and $W_0^{\lambda2600}>0.5\angmath$ yielded
DLAs.  Accordingly \citet{rt2006} preferentially observed systems with
both strong \mgii and \feiinsp.  Notably, the only $\mgiiwr<0.6\angmath$
systems included in this sample were those already observed in
\citet{rt2000}, or those serendipitously along the same QSO sightlines
as other systems with strong \mgii and \feiinsp.

Figure \ref{fig:rtsub} shows the cumulative distribution functions for
$\mgiiwr$ (upper panel) and $W_0^{\lambda2600}$ (lower panel) for the
full low redshift $z<2$ sample described in Section
\ref{sec:ions_subsample} (black lines) and the \emph{HST} \hi sample (light gray
lines).  We derived the CDFs for \feii 2600, which include upper
limits, from the Kaplan-Meier estimator using the Astronomy Survival
Analysis (ASURV) package \citep{asurv}, which implements the methods
of \citet{unisurvival}.  The CDFs show how, by construction, the \emph{HST}
sample systematically favors high rest equivalent width systems in
both \mgii and \feiinsp, relative to a randomly selected population of
intervening absorbers.

\begin{figure}
\epsscale{1.20}
\advplotone{Ions}{rtsub}{5.5in}{4.25in}
\caption{Cumulative distribution functions in rest-frame equivalent
  width for \mgii 2796 (upper panel) and \feii 2600 (lower panel).
  The black lines are for the low redshift sample ($0<z<2$,
  $\mgiiwr\geq0.3\angmath$) described in Section \ref{sec:ions_subsample},
  and the light gray lines are for the \emph{HST}
  sample of \citet{rt2006}.  The \emph{HST} sample has an overabundance of
  strong \mgii 2796 and \feii 2600 absorbers relative to the low
  redshift sample.  The CDFs for the subsample of \nrtsub~ \emph{HST} systems
  chosen to match the low redshift sample's \mgii 2796 and \feii 2600
  distributions are overplotted with dark gray lines.  A few
  representative error bars have been overplotted.}
\label{fig:rtsub}
\end{figure}

For statistical analysis, we therefore extracted a subset of the \emph{HST}
sample to match the $\mgiiwr$ and $W_0^{\lambda2600}$ distributions of
the low redshift sample.  First, we made an estimate of the number of
\emph{HST} systems required in each of \rtnbins~ logarithmically-spaced
$\mgiiwr$ bins in order to match the distribution of the low redshift
sample.  We then ran a Monte Carlo simulation that provided \rttmax~
possible realizations with this broad-stroke binning property.  For
each of these realizations, we calculated the two-sample logrank,
Gehan, and Peto-Prentice probabilities that the low redshift and \emph{HST}
samples were drawn from the same distribution for {\em both} the \mgii
and \feii distributions using the ASURV package.  The final subsample
exhibited the highest geometric mean of these six probabilities.

The CDFs for this final subsample are shown as the dark gray lines in
Figure \ref{fig:rtsub}.  The final subsample contains \nrtsub~ systems
ranging in redshift from \rtsubzmin~ to \rtsubzmax, with a mean
redshift of $\bar{z}=$\rtsubzmean.  The geometric mean of the three
two-sample tests performed is \rtmgeoprob\% for the $\mgiiwr$
distributions and \rtfgeoprob\% for $W_0^{\lambda2600}$ distributions.
Clearly this procedure has effectively eliminated the bias towards
stronger \mgii and \feii systems.

Our high redshift \mgiinsp-selected $\nhi$ sample should not suffer from
similar selection bias since the \mgii systems were selected randomly
and \hi measurements were obtained for all systems not blocked by an
intervening Lyman limit.  To verify this expectation, we performed a
Kolmogorov-Smirnov test, finding a \hiznhiks\% probability that the
$\mgiiwr$ distribution for our \hinsp-measured subset is drawn from the
same distribution as the full high redshift sample.

\subsubsection{Normalizing biases in the Metallicity Sample}
\label{sec:ions_prsub}

As accounted by the authors, the \citet{prochaska_dla} sample of DLA
abundance measurements represents an inclusive compilation of observed
DLAs at the time of publication, but it is not a statistically
characterized random sample (mostly likely due to observer selection
bias towards stronger $\nhi$ systems).  This is evident in Figure
\ref{fig:prsub}, where the $\nhi$ CDF for this sample (light gray
line) is shown against that of a DLA survey from the SDSS DR5, as provided by 
\citet[][black line; taken to be a statistically random
sample]{sdss_dla_dr5}.  Both samples in this plot are limited to $z\geq2$ and
$\lognhi\geq20.3$ \pcmsq~systems, for which the \citet{prochaska_dla}
sample contains \npsub~ systems and the DR5 sample contains \ndsub~
systems.

\begin{figure}
\epsscale{1.20}
\advplotone{Ions}{prsub}{5.5in}{4.25in}
\caption{Cumulative distribution functions in $\nhi$ for
  systems in the \citet{prochaska_dla} sample (light gray line) and
  the SDSS DR5 DLA sample (black line) with $z\geq2$ and
  $\lognhi\geq20.3$~\pcmsq~ (with a few representative error bars
  overplotted).  The \citet{prochaska_dla} sample shows a preference
  for stronger $\nhi$ systems relative to the SDSS survey, possibly a
  sign of observer bias in selecting targets.  The dark gray line
  represents the unbiased \citet{prochaska_dla} DLA subsample with
  \ndla~ systems derived using the process described in Section
  \ref{sec:ions_prsub}.}
\label{fig:prsub}
\end{figure}

We created our final high redshift metallicity comparison set from the
\citet{prochaska_dla} systems using the SDSS DR5 DLA sample as a
comparison set and a procedure similar to that used in Section
\ref{sec:ions_rtsub}.  Our Monte Carlo simulation generated \prtmax~
possible comparison samples that met the broad-scale binning
requirements in $\lognhi$ set by the SDSS DR5 DLA $z\geq2$ sample.
For each of these realizations, we calculated the two-sample logrank
and Gehan probabilities that the sample was drawn from the same
distribution as the SDSS DR5 DLA sample. (The Peto-Prentice test is redundant with the Gehan test in the absence of upper limits, and was therefore excluded.)  We chose the final sample to
be that with the highest geometric mean of these two probabilities.

The final high redshift metallicity sample consists of \ndla~ DLAs,
ranging in redshift from \psubdlazmin~ to \psubdlazmax, with a mean
redshift of $\bar{z}=$\psubzmean.  This distribution and the SDSS
$z\geq2$ DR5 DLA distribution had two-sample logrank and Gehan
probabilities of \dlalogrank\% and \dlagehhyp\% of being drawn from
the same parent distribution.  The CDF of the final DLA/abundance
control sample is shown in Figure \ref{fig:prsub} as the dark gray
line.

\section{Results}
\label{sec:ions_results}

\begin{figure*}
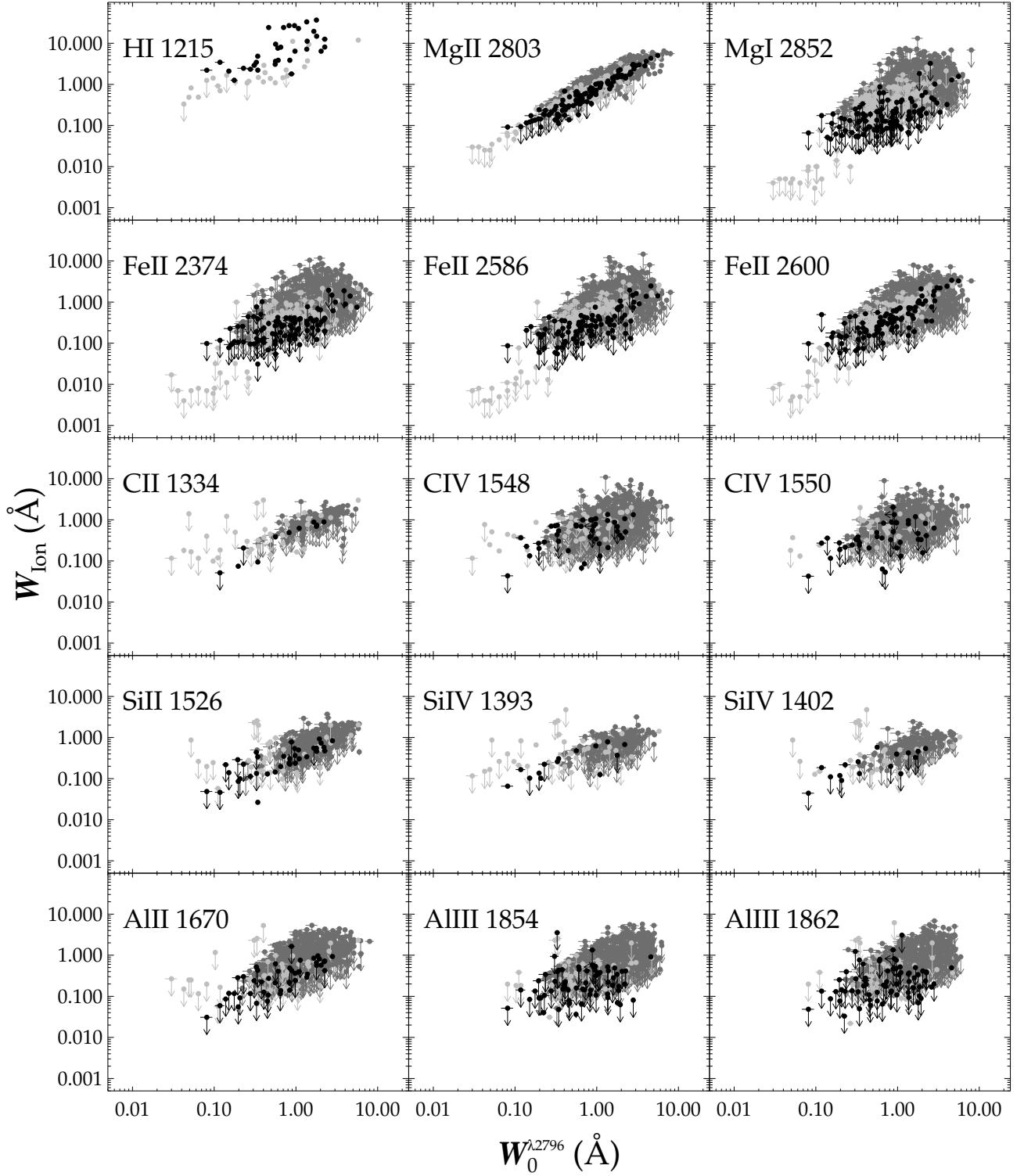

\epsscale{1.15}
\advplotone{Ions}{ion_comparison}{6.25in}{6.90in}
\caption{Rest-frame equivalent widths for various ions versus
  $\mgiiwr$.  The dark gray points are from the full SDSS DR7 sample
  ($0.36 < z < 2.23$), and the light gray points and black points are
  the low ($0<z<2$) and high ($z\geq2$) redshift systems, respectively, for the full
  literature compilation and FIRE sets (not the completeness-corrected
  samples).  All of the FIRE measurements are listed in Tables
  \ref{tab:mglist}, \ref{tab:olist}, \ref{tab:felist}, \ref{tab:silist}, and
  \ref{tab:clist}.  A subset of the literature
  compilation measurements are listed in Table \ref{tab:lowzsample}.}
\label{fig:ion_comparison}
\end{figure*}

Figure \ref{fig:ion_comparison} displays scatter plots of all metal
rest equivalent widths from the low redshift literature compilation, FIRE, 
and SDSS DR7 samples (not just those included in the unbiased subsamples).  
The black and light gray dots represent systems with $z\geq2$ and $z<2$, respectively, 
from the FIRE and literature compilation samples, and the dark gray dots 
represent measurements from the SDSS DR7 sample.

The scatter plots for individual transitions indicate that for most
elements, the loci occupied by low and high redshift points are
largely overlapping.  The major exception is for \hinsp, which is clearly
higher for the high redshift sample.  There are hints of offsets in
\aliii and select transitions of \feiinsp.  But the influence of upper
limits and saturation are not at first clear in this view.  To better
quantify these effects, we have therefore constructed CDFs (accounting
for upper limits) for each ion ratio relative to \mgii and performed
two-sample tests to discern whether evolutionary trends may be
extracted from the scatter.

\subsection{\hi Evolution in \mgiinsp-selected Systems}
\label{sec:ions_hiresults}

\begin{figure}
\epsscale{1.20}
\advplotone{Ions}{compare_nhi_samples}{5.50in}{4.0in}
\caption{Cumulative distribution functions for $\nhi$ for the
  unbiased, $\mgiiwr\geq0.3\angmath$ low redshift ($z<2$; thick, light
  gray line) and high redshift ($z\geq2$; thick, black line) samples
  described in Section \ref{sec:ions_rtsub} (overplotted with a few
  representative error bars).  The \mgii absorbers from the high
  redshift sample are typically associated with much larger \hi column
  densities.  A K-S test provided only a \nhiks\% probability that the
  two samples were drawn from the same distribution.  Also overplotted
  are the 2nd, 3rd, and 4th best-matched low redshift $\nhi$ samples
  (thin, light gray lines) from the MC simulation discussed in Section
  \ref{sec:ions_rtsub}.  These also provide K-S probabilities of $<0.1$\%,
  suggesting that this result is robust to the exact sample chosen.  The thin, black line 
  represents the high redshift sample, but with $\lognhi$ values $3\sigma$ lower 
  than calculated.  This distribution has only a \nhisyserrprob\% K-S probability 
  of deriving from the same parent distribution as the low redshift sample, 
  suggesting that our results are robust to large, systematic overestimates of 
  $\lognhi$ as well.}
\label{fig:compare_nhi_samples}
\end{figure}

Figure \ref{fig:compare_nhi_samples} displays the CDFs of $\nhi$ for
the $\mgiiwr\geq0.3\angmath$ \mgiinsp-selected systems in the low ($z<2$; thick, 
light gray line) and high ($z\geq2$; thick, black line) redshift samples.
Every such \mgii system at $z\geq2$ is optically thick to \hi, with
\ndlasfire~ of \nnhifire~ exhibiting DLA column densities
($\perdlasfire^{+\perdlashigh}_{-\perdlaslow}$\%).  The \mgiinsp-selected sample at $z>2$ has a mean log column density (\nmeanfiresub~
\pcmsq) that
nearly meets the DLA threshold.

From this plot, it is also clear that \mgii absorbers are associated
with stronger $\nhi$ absorption at high redshifts.  In spite of the
relatively small sample sizes, a Kolmogorov-Smirnov test gives only a
\nhiks~ percent probability that the two samples were drawn from the
same distribution.  Also overplotted (thin, light gray lines) are the 2nd, 3rd, and 4th best-matched low redshift $\nhi$ samples from the MC simulation discussed in Section \ref{sec:ions_rtsub}.  These also provide K-S probabilities of $<0.1$\%, suggesting that this result is robust to the exact sample chosen.  In addition, the overplotted thin, black line represents the CDF for the high redshift sample, but with all measurements shifted $3\sigma$ lower than calculated.  This distribution has only a \nhisyserrprob\% K-S probability of deriving from the same parent distribution as the low redshift sample, suggesting that our results are also robust to large, systematic overestimates of $\lognhi$.

Only $\rtperdlas^{+\rtperdlashigh}_{-\rtperdlaslow}$\% of the low redshift sample systems are associated with DLAs, as compared to $\perdlasfire^{+\perdlashigh}_{-\perdlaslow}$\%, or about than 2 in 5, in the high redshift sample.  Figure \ref{fig:perdlas} shows the percentage of \mgii systems in the low (light gray line) and high (black line) redshift samples exhibiting DLA column densities for $0.3\angmath\leq\mgiiwr\leq W_{\hbox{0,max}}$ and varying values of $W_{\hbox{0,max}}$.  

\begin{figure}
\epsscale{1.20}
\advplotone{Ions}{perdlas}{5.5in}{4.00in}
\caption{Percentage of \mgii absorbers with strengths in the range
  $0.3\angmath\leq \mgiiwr \leq W_{0,\hbox{max}}$ associated with DLAs
  for the low ($z<2$; light gray) and high ($z\geq2$; black) redshift
  samples described in Section \ref{sec:ions_rtsub} (with a few
  representative error bars overplotted).  The DLA percentage at large
  $\mgiiwr$ is significantly higher for the high redshift sample
  ($\perdlasfire^{+\perdlashigh}_{-\perdlaslow}$\%) than for the low
  redshift sample
  ($\rtperdlas^{+\rtperdlashigh}_{-\rtperdlaslow}$\%).}
\label{fig:perdlas}
\end{figure}

Figure \ref{fig:z_nhi} more directly illustrates the redshift
evolution in $\nhi$ by plotting the \hi column densities for all
systems in the representative subsets against redshift.  The solid
line represents an interative sigma-clipped linear fit $\lognhi =
(\znhim\pm\znhimerr) z + (\znhib\pm\znhiberr)$ \pcmsq.  The dotted
lines are the one sigma limits.  The evolution is significant at a
$>4\sigma$ level, with the best fit line increasing from
$\lognhi\sim19$ at $z\sim0$ to above the DLA threshold (dashed line)
for $z>4$.

\begin{figure}
\epsscale{1.20}
\advplotone{Ions}{z_nhi}{5.50in}{4.00in}
\caption{$\nhi$ column density as a function of absorber redshift for
  \mgiinsp-selected systems for the \emph{HST} sample (light gray) and
  high redshift FIRE sample (black) described in Section
  \ref{sec:ions_rtsub}.  The horizontal dashed line is at the damped
  \lya cutoff, $\nhi=$2e20 \pcmsq.  The higher redshift sample
  contains higher \hi column densities and more DLAs, consistent with
  the results of Figures \ref{fig:compare_nhi_samples} and
  \ref{fig:perdlas}.  The solid line is the sigma-clipped linear fit
  $\lognhi = (\znhim\pm\znhimerr) z + (\znhib\pm\znhiberr)$ \pcmsq.
  The dotted lines are the one sigma limits.}
\label{fig:z_nhi}
\end{figure}

\subsection{Chemical Evolution in \mgiinsp-selected Systems}
\label{sec:ions_chemresults}

Figure \ref{fig:cdf_ratios} provides CDFs---calculated using the
Kaplan-Meier estimator---of the equivalent width for each heavy
element ion we measured, after normalizing by $\mgiiwr$.  Separate
curves are shown for the low redshift sample ($z<2$; light gray
lines), the high redshift sample ($z\geq2$; black lines), and the SDSS
sample ($0.36<z<2$; dark gray lines).  Table \ref{tab:asurv2} lists the
sample median for each ratio considered.  For \hinsp, $\hiwr/\mgiiwr$ values 
derived from the \emph{HST} sample of Section \ref{sec:ions_rtsub} (using a curve of growth 
analysis with $b=30$ \kms) are substituted for the SDSS DR7 sample (which
contains no $\hiwr$ measurements).  For each transition, we performed two-sample 
tests to assess the probability that the low and high redshift
CDFs derive from a common parent population.  We used three separate
tests, which each account for upper limits, to generate these
probabilities (stored in Table \ref{tab:asurv2}): the logrank $P_{LR}$, Gehan $P_{G}$, and Peto-Prentice
$P_{PP}$ tests.

Figure \ref{fig:cdf_ratios3} and Table \ref{tab:asurv3} provide the analogous 
CDFs, median ratios, and two-sample test results with three redshift bins instead of two.  In Figure
\ref{fig:cdf_ratios3}, the low and high redshift samples (now at
$0<z<1.5$ and $z\geq3$) are still drawn in light gray and black,
respectively, with a new mid-range redshift sample ($1.5\leq z<3$)
depicted in dark gray.  (No SDSS data are included in this plot).  In
Table \ref{tab:asurv3}, the $L$, $M$, and $H$ subscripts on the two-sample 
probabilities denote which two of the low, mid-range, and high
redshift samples were used in the calculation.

\begin{figure*}
\epsscale{1.15}
\advplotone{Ions}{cdf_ratios}{6.00in}{4.65in}
\caption{Kaplan-Meier cumulative distribution functions for the ratios
  of the rest equivalent widths of various ions with $\mgiiwr$ for the
  low redshift ($0<z<2$; light gray), high redshift ($z\geq2$; black),
  and SDSS DR7 ($0.36<z<2$; dark gray) samples described in Section
  \ref{sec:ions_subsample}.  For \hi1215, the \emph{HST} sample from
  Section \ref{sec:ions_rtsub} (with $\hiwr$ from a curve of growth
  analysis with $b=30$ \kms) is used in place of the SDSS sample
  (which contains no $W_0^{\lambda1215}$ measurements).  Detailed
  statistics associated with this figure and the sub-samples are given
  in Tables \ref{tab:sample_counts} and \ref{tab:asurv2}. A few
  representative error bars have been overplotted.  While the CDFs for
  a few of the ions (\mginsp, \aliiinsp\dots) show signs of evolution,
  only the ratio of $\hiwr/\mgiiwr$ shows clear signs of strong
  evolution.}
\label{fig:cdf_ratios}
\end{figure*}

\begin{figure*}
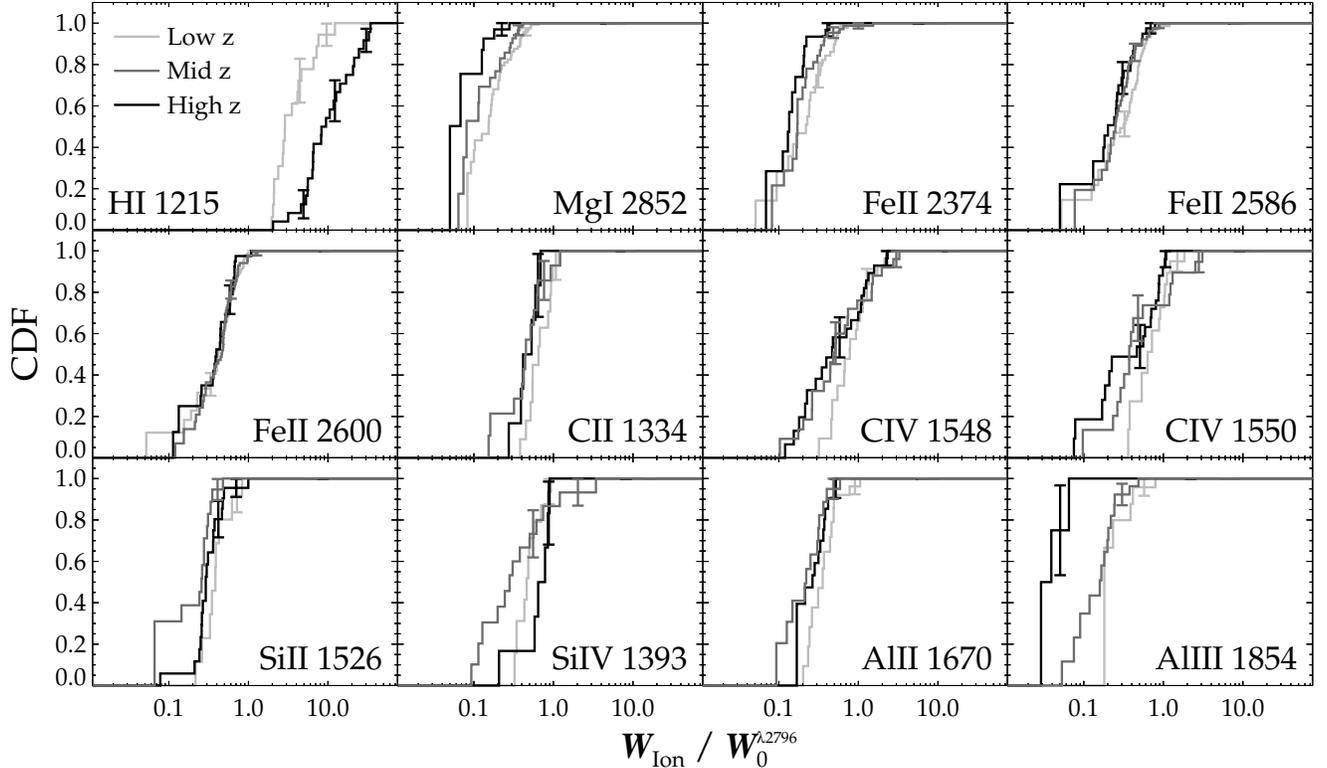

\epsscale{1.15}
\advplotone{Ions}{cdf_ratios3}{6.00in}{4.65in}
\caption{Analogous plot to Figure \ref{fig:cdf_ratios}, but for the
  three redshift-binned low redshift ($0<z\leq1.5$; light gray),
  medium redshift ($1.5\leq z<3$; dark gray) and high redshift
  ($z\geq3$; black) samples described in Section \ref{sec:ions_subsample}.
  (The medium redshift sample contained only a few $\hiwr$ measurements and is
  excluded from that panel).  Detailed statistics associated with this figure and the sub-samples are given in Tables \ref{tab:sample_counts3} and \ref{tab:asurv3}.  A few representative error bars have been overplotted.  The evolution of \civ appears strongest at $z<1.5$.}
\label{fig:cdf_ratios3}
\end{figure*}

The \hi equivalent width distributions confirm the trend seen in
Figure \ref{fig:compare_nhi_samples}, that the high redshift systems have
markedly more neutral hydrogen for a given \mgii absorption strength.
The mean ratio $\hiwr/\mgiiwr$ increases from
$\cdfhimeanlow$ in the low redshift sample to $\cdfhimeanhigh$ at high
redshift.  All three two-sample tests yield a $<1\%$ probability
that the low and high redshift distributions are alike.  Figure \ref{fig:z_hiratios}, a scatter 
plot of $\hiwr/\mgiiwr$ versus redshift for the representative low and high 
redshift samples, highlights this evolution.  The solid line is the robust 
linear fit $W_0^{\lambda1215}/\mgiiwr = (\zratm\pm\zratmerr) z + (\zratb\pm\zratberr)$.  The dotted 
lines are one sigma limits.

\begin{figure}
\epsscale{1.20}
\advplotone{Ions}{z_hiratios}{5.5in}{4.00in}
\caption{Relative absorption strength $W_0^{\lambda1215}/\mgiiwr$ as a
  function of absorber redshift for the low redshift (light gray) and
  high redshift (black) samples described in Section
  \ref{sec:ions_subsample}.  The higher redshift sample contains more \hi
  absorption relative to \mgiinsp, consistent with the results of Figure
  \ref{fig:cdf_ratios}.  The solid line is the robust linear fit
  $W_0^{\lambda1215}/\mgiiwr = (\zratm\pm\zratmerr) z +
  (\zratb\pm\zratberr)$.  The dotted lines are the one sigma limits.}
\label{fig:z_hiratios}
\end{figure}

For heavy element transitions, however, the difference between the
high and low redshift samples is much less pronounced.  For the 
singly ionized species in particular (see \ciinsp, \aliinsp,
\siiinsp) the CDFs are nearly indistinguishable,
and the two-sample tests often produce high probabilities of a draw from the
same parent population ($>50-95$\%) and never produce low $<10\%$ probabilities.
  The median values of
\alii and \siii only vary by 0.01-0.03 in ratio.  Those of \cii vary by slightly 
more (0.08), but this ion suffers from considerably fewer counts.

For \feiinsp, the samples are also very similar, though the exact degree
depends on the multiplet transition used.  The 1608, 2344, 2382, and
2600\AA ~ lines show no statistically significant difference between
low and high redshift, while the 2374 and 2586 appear statistically
smaller (by 0.05-0.1 in ratio) at high redshift.  The latter two
lines show a $\lesssim 5$\% chance of deriving from the same parent
population at high $z\geq2$ and low $z<2$ redshifts.  

Although not directly testable at our data's resolution, this
difference could naturally arise from a combination of line saturation
and small number counts.  Saturation would affect the 2344, 2382, and
2600\AA ~transitions because of their large oscillator strengths, so
even a substantial change in $N(\hbox{\feiinsp})$ would yield little change
in equivalent width, particularly if the velocity spread is similar to
that of \mgii (which we normalize out by taking the equivalent width
ratio).  The lower oscillator strengths of the 2374 and 2586 lines may
leave them unsaturated, increasing their sensitivity to evolution.  The
1608\AA ~line does not fit into this story as its oscillator strength
is also low, but our statistics on this transition are relatively poor
compared to the redder transitions, so the significance is less
strong.

In contrast to the singly ionized species, \civnsp, \siiv and especially \aliii
do appear to evolve, in the sense that the highly ionized lines are
weaker toward high redshift.  For a given $\mgiiwr$, the median \aliii
line ratio $W_0^{\lambda1854}/\mgiiwr$ decreases from
$\cdfaliiiameanlow$ at low redshift to
$\cdfaliiiameanhigh$ at high redshift for the two bin samples.  All
three two-sample distribution tests suggest a very small
$\lesssim0.1\%$ probability of no evolution between the highest redshift set and 
the other two.

Likewise the \civ ratio
is reduced toward higher redshift, with the 1550\AA~ component showing
a more statistically significant change (again, possibly a saturation
effect) and low $\lesssim10\%$ probabilities are being drawn from the same distribution.  
This \civ evolution is most pronounced when dividing the sample 
into three redshift bins.  With these divisions, we find very high $40-60\%$ probabilities 
that the \civ1550 mid and high redshift distributions are the same, suggesting 
that the most significant changes occur at $z<1.5$.  

Among the multiply ionized species, \siiv alone seems not to evolve:
the two-sample tests yield probabilities of $\sim35-65\%$ that the
high and low redshift samples are drawn from the same distribution,
although this could be a result of low counts.

Finally, the $W_0^{\lambda2852}/\mgiiwr$ ratio between \mgi and \mgii
decreases in a statistically significant manner as redshift increases
for the two redshift sample scenario, with all three two-sample
distribution tests giving a $\lesssim0.1$\% probability that the
ratios are drawn from the same distribution.  The SDSS DR7 sample confirms the low redshift sample's relative strength in \mgi absorption. The CDFs and analogous
two-sample tests with three redshift bins suggest that the strongest
evolution occurred before $z=3$.  The downward evolution in \mgi is
slightly surprising given the basically unchanging nature of the other
low-ionization lines.

\subsection{Chemical Composition in Weak and Strong \mgiinsp-selected Systems}
\label{sec:ions_chemcompresults}

Figure \ref{fig:cdf_ratios_wr} gives the Kaplan-Meier derived CDFs for
various ions relative to \mgiinsp, but now divided into two samples by
$\mgiiwr$ rather than redshift.  The light gray line represents weak systems
($0.3\angmath\leq\mgiiwr\leq1.0\angmath$), while the black line
represents strong systems ($\mgiiwr>1.0\angmath$).  Table
\ref{tab:asurvwr} provides the median ratios and logrank $P_{LR}$,
Gehan $P_G$, and Peto-Prentice $P_{PP}$ two-sample test probabilities for each ion. 
Figure \ref{fig:cdf_ratios_wrz} and Table \ref{tab:asurvwrz} provide the analogous information 
as Figure \ref{fig:cdf_ratios_wr} and Table \ref{tab:asurvwr}, respectively, but now with the weak and 
strong absorber classes further divided by redshift as well.  In Figure \ref{fig:cdf_ratios_wrz}, the 
dotted and solid lines represent weak and strong absorbers, respectively, and the light gray and black 
lines represent low ($z<2$) and high ($z\geq2$) redshifts, respectively.

\begin{figure*}
\epsscale{1.15}
\advplotone{Ions}{cdf_ratios_wr}{6.00in}{4.65in}
\caption{Analogous plot to Figure \ref{fig:cdf_ratios}, but for the
  weak ($0.3\leq \mgiiwr\leq1.0$; light gray) and strong
  ($\mgiiwr>1.0$; black) \mgii absorption samples described in Section
  \ref{sec:ions_wrsubsample}.  Detailed statistics associated with this figure and the sub-samples are given in Tables \ref{tab:sample_counts_wr} and \ref{tab:asurvwr}.  A few representative error
  bars have been overplotted.  The samples appear to contain the same
  amount of \hi \emph{relative to} \mgiinsp, but the strong absorbers
  contain less \civ and \siivnsp, perhaps because of their larger
  absolute amounts of \hi shielding these ions.}
\label{fig:cdf_ratios_wr}
\end{figure*}

\begin{figure*}
\epsscale{1.15}
\advplotone{Ions}{cdf_ratios_wrz}{6.00in}{4.65in}
\caption{Analogous plot to Figure \ref{fig:cdf_ratios_wrz}, but with
  the weak ($0.3\leq \mgiiwr\leq1.0$; dotted lines) and strong
  ($\mgiiwr>1.0$; solid lines) \mgii absorption samples divided into
  low ($z<2$; gray lines) and high ($z\geq2$; black lines) redshift parts
  as well.  Detailed statistics associated with this figure and the sub-samples are given in Tables \ref{tab:sample_counts_wrz} and \ref{tab:asurvwrz}.  A few representative error
  bars have been overplotted.  The distribution of $\hiwr/\mgiiwr$
  appears the same between the weak and strong samples for both the
  low and high redshift cuts.}
\label{fig:cdf_ratios_wrz}
\end{figure*}

One noticeable feature is that the \lya
CDF appears identical for the strong and weak \mgii samples:  The
median ratio of $W_0^{\lambda1215}/\mgiiwr$ is higher
for the weak \mgii absorbers
($\cdfwhimeanlow$ vs. $\cdfwhimeanhigh$), but the two-sample
tests find any differences to be statistically insignificant.  
Moreover, Figure \ref{fig:cdf_ratios_wrz} shows that both weak and strong \mgii 
systems evolve very similarly (both strongly) with redshift

In contrast to \hinsp, the two-sample tests all suggest a very small probability
($\lesssim5\%$) that \mgi 2852 absorption is the same between weak and strong \mgii systems, with a
tendency towards more relative amounts of \mgi in stronger \mgii
systems as expected.  The other low ionization species' (\ciinsp, \feiinsp, \siiinsp, and \aliinsp) 
ratios display much greater similarity between weak and strong \mgii systems, and none 
of their CDFs show significant qualitative differences.  Although some of these
distributions possess statistically significant differences (e.g., \feii 2600, which has 
probabilities $\lesssim1\%$ of having ratios with weak and strong absorbers drawn from the 
same distribution), the evolution detected even in these cases is quite weak.

The higher ionization lines \civ and \siiv, however, decrease significantly in
strength as $\mgiiwr$ increases ($\cdfwcivameanlow$ to $\cdfwcivameanhigh$ for \civ1548, for example).  
The two-sample tests suggest a very small probability that these samples are drawn from the same 
distributions for weak and strong absorbers ($\lesssim3$\%).  The may result from \hinsp-shielding, since 
the strong absorbers have larger \emph{absolute} amounts of \hinsp, even though their relative 
amounts are nearly identical.  The strong \mgii systems have 
low $\lesssim4\%$ two-sample probabilities that their low and high redshift subsets have $\civwr/\mgiiwr$ and $W_0^{\lambda1550}/\mgiiwr$ ratios
drawn from the same distribution, while the weak \mgii systems have high $\sim30-75\%$ probabilities 
of no redshift evolution.  Interestingly, the \siiv distributions hint at the reverse, although 
the evidence for redshift evolution for the weak absorbers is not statistically strong.  
Not much may be deduced from the \aliii 
distributions since this ion suffers from small counts among weak \mgii systems.

\subsection{Metallicity of \mgiinsp-selected Systems at High Redshift}
\label{sec:ions_metals}

Figure \ref{fig:metals} gives the Fe, C, Si, and Al metallicities for
the \mgiinsp-selected high redshift FIRE sample (black points) and the
general high redshift \citet{prochaska_dla} metallicity subsample from Section \ref{sec:ions_prsub} (gray
points).  Triangular points for the FIRE data denote weak \mgii
absorbers ($0.3\angmath\leq \mgiiwr\leq1.0\angmath$) and the diamonds
denote strong absorbers ($\mgiiwr>1.0\angmath$).  Table
\ref{tab:nhi_table} provides these metallicity measurements for the
FIRE systems.

As previously stated, we have employed very conservative criteria for
flagging saturated lines in our moderate resolution spectra.  This
explains why the black points are mostly lower limits in these
metallicity measurements.  Indeed all of the measurements made for
DLAs \citep[which may be directly compared to][]{prochaska_dla} are
lower limits.  Despite this, the data still establish that
\mgiinsp-selected systems are not metal poor with respect to the general
DLA population.  (The possibility that \mgii absorbers are actually probing the full, underlying DLA population is addressed later in the Discussion, Section \ref{sec:ions_dlas}.)

In fact, the limits imply quite high abundances of 0.1 Solar or more
for weak \mgii systems, which are predominantly optically thick
sub-DLA absorbers as viewed in \hinsp.  Some caution is warranted for
these points since we have not included ionization corrections.
Detailed study of the ionization in $z\sim 3$ sub-DLAs by
\citet{peroux2007} suggests that such ionization corrections generally
decrease the resulting metallicity.  The magnitude of the effect
depends on $\nhi$, with systems at $\lognhi\sim19$~\pcmsq~ requiring
a $\sim 0.3$ dex correction, and stronger systems requiring less until
the DLA threshold is reached.  Still, these limits would still fall
near 10\% solar for many of the \mgii sub-DLAs, which (like most
sub-DLAs) appear to be much more metal rich than the IGM and may in
fact be more enriched than classical DLAs.

\begin{figure*}
\epsscale{1.15}
\advplotone{Ions}{metals}{6in}{4.75in}
\caption{Metallicities (relative to solar) for the \mgiinsp-selected
  (black) and non-\mgiinsp-selected (light gray) high redshift samples
  ($z\geq2$) described in Section \ref{sec:ions_prsub}.  For \mgiinsp-selected
  systems, weak \mgii absorbers ($0.3\angmath\leq
  \mgiiwr\leq1.0\angmath$) are shown as triangles and strong
  ($\mgiiwr>1.0\angmath$) as diamonds.  The dashed vertical line is at
  the DLA cut, $\nhi=$2e20 \pcmsq.  All metallicities for
  \mgiinsp-selected DLA systems are lower limits, making comparison
  difficult, but the \mgiinsp-selected systems are not metal-poor
  compared to the general population of absorbers at these redshifts.
  No ionization corrections have been applied, potentially leading to
  overestimates of up to $\sim 0.3$ dex for the lower $\nhi$ systems'
  metallicities in this plot \citep{peroux2007}.  Even with this
  correction, the lower limits of the lower $\nhi$ systems approach a
  tenth of solar.}
\label{fig:metals}
\end{figure*}

\section{Discussion}
\label{sec:ions_discuss}

\subsection{Taxonomy and Evolution of Classes}
\label{sec:ions_classify}

Numerous studies in the literature have proposed that \mgii traces
multiple physical environments.  These are variously based on
differential evolution in $dN/dX$ (\paperone), statistical studies of
\mgii host galaxy color \citep{zibetti2007colors,lund2009},
galaxy-\mgii clustering analysis
\citep{bouche2006anticorrelation,gauthier,nestor2010strong}, and
studies of galaxy-absorber projected inclination \citep{bordoloi,kacprzak2011incl}.
  Some of these studies suggest that stronger \mgii
systems are found near star forming galaxies and may be related to
outflows.  Indeed, models deriving strong $\mgiiwr>1\angmath$ absorption from star-forming disks and their associated outflowing interstellar material \citep[e.g., ][]{chelouche2010} show better agreement with the empirically measured $z>2.5$ $dN/dX$ than halo occupation models \citep[e.g., ][]{chenandtinker2009}.  However the connection between strong \mgii absorption and star formation is not universally found, and 
the use of $\mgiiwr$ alone to distinguish outflowing from accreting
\mgii absorbers is almost certainly an oversimplification.

This motivates us to explore other schemes for classifying \mgii
absorbers, since we have access to numerous high- and low-ionization
transitions.  We have adopted the methodology of
\citet{churchill_zoo}, who developed a classification taxonomy for
\mgii absorbers based on a multivariate clustering analysis for \mgii
systems at $z=0.4-1.4$.  The analysis incorporates measurements of
equivalent width for \mgiinsp, \hinsp, \feiinsp, and \civnsp, as well as the
kinematic spread for \mgii (denoted $\mgiiks$).

\citet{churchill_zoo} contains details of the methodology.  Briefly,
one must first ``standardize'' the distributions for these five
properties into an $N(0,1)$ Gaussian form, and then implement a
$K$-means clustering algorithm that moves systems between clusters
until the variability within clusters is minimized and across clusters
is maximized.  The 45 \mgii systems grouped in this way segregated into five
statistically distinct classes.

\citet{churchill_zoo} named the five classes as follows: 
\begin{enumerate}
\item{\emph{classic systems} (24\%), which have $\mgiiwr$, $\feiiwr$,
  $\civwr$, $\hiwr$, and $\mgiiks$ within 0.5$\sigma$ of the overall
  normalized sample mean.}
\item{\civnsp-\emph{deficient systems} (18\%), which are otherwise
  identical to classic systems but have significantly less $\civwr$:  the mean in standardized units is more than 1.5 less.}
\item{ \emph{DLA/\hinsp-rich systems} (13\%), which have stronger $\mgiiwr$ and much
  stronger $\feiiwr$ and $\hiwr$ than classic systems, but similar
  $\mgiiks$ and weaker $\civwr$.}
\item{\emph{double systems} (7\%), which have larger equivalent width
  and velocity spread than classic systems, including much stronger
  ($>2\times$) $\mgiiwr$, $\civwr$, and $\mgiiks$.  The naming convention
  for this class was inspired by the work of \citet{bond}, who
  identified such systems as double-troughed absorbers in HIRES
  spectra.}
\item{\emph{single/weak systems} (38\%), which are single component,
  narrow lines with the means of $\mgiiwr$ and $\civwr$ in standardized units weaker by $\sim1$ compared to classic  systems.}
\end{enumerate}

For reasons of completeness in the FIRE sample, we limit our
discussion to systems with $\mgiiwr>0.3$\AA, effectively eliminating
the single/weak systems from consideration.  This leaves four
classification bins for the high redshift systems.

We first explored direct application of Churchill's method using the
low redshift standardization parameters, to see how the population
evolves relative to an absolute benchmark.  This exercise was less
illuminating than anticipated:  the typical
system at high redshift has slightly weaker \mgii (see Figure 15 in \paperonesp 
regarding the evolution of the typical system size $W_*$), which would imply a
classic or weak classification, but Figure
\ref{fig:compare_nhi_samples} shows that $\nhi$~ in the corresponding
absorbers is higher, suggesting a DLA or double classification.  In
other words, the high redshift systems would require a separate class
altogether which possesses less heavy elements than the low redshift classes,
but contains more \hi absorption.

Next, we attempted a new classification where each absorber was
standardized to the properties of the typical system at its respective
redshift.  This requires a slightly different interpretation but
produces a more well-defined taxonomy.

We produced standardization distributions for three redshift bins
($z<3$, $3\leq z<4$, and $z\geq4$), using the high redshift sample 
 of Section \ref{sec:ions_subsample}.  The actual
standardization, which maps the observed CDF onto a standard normal distribution,
 is accomplished using
\begin{equation}
y_{i}=\left\{
\begin{array}{ll}
-\sqrt{2} \hbox{erfc}^{-1}\left(P(x_i)\right) & \hspace{1em}P(x_i) \leq 0.5\\
\sqrt{2} \hbox{erf}^{-1}\left(2P(x_i)-1\right) & \hspace{1em}P(x_i) > 0.5,
\end{array}\right.
\label{eq:std}
\end{equation}
where $x_i$ is the original absorption value, $y_i$ is the
standardized value, $P(x)$ is the CDF, erf$^{-1}(x)$ and
erfc$^{-1}(x)$ are the inverse error and complementary error functions, and
the index $i$ indicates each system considered.  This exercise is
repeated for each redshift bin and absorption property used for
classification.
Since our highest redshift bin ($z\geq4$) contains only one measured
value for $\hiwr$, we pooled the $\hiwr$ values from the two largest
redshift bins when calculating their $\hiwr$ CDFs.

A full treatment would then require re-calculation of the $K$-means
clustering algorithm and generation of new classes for each redshift
bin.  But this is not practical for the high redshift sample because
for many systems we can only measure 2 or 3 of the 5 classification
observables.  This is partly a consequence of the QSO sightline
selection for the FIRE survey, for which we prioritized high redshift
objects to maximize pathlength and \mgii sample size, thereby
minimizing $dN/dX$ errors at $z>3$.  While accomplishing these goals,
the FIRE sample is not ideally suited for a $z>2$ classification
analysis.  In particular, by choosing QSOs at high emission redshift
one increases the likelihood that \civ and \hi measurements at
$z\sim2-2.5$ will be lost due to absorption from the \lya forest
and/or higher redshift Lyman limit systems.  In practice, the lowest
redshift for which we have $\civwr$ and $\hiwr$ measurements in the
FIRE statistical sample are $z=2.749$ and $z=2.593$, respectively.
The ideal classification sample would have contained more background
objects at $z_{QSO}=2.5-3$ to avoid this paucity of $\civwr$ and
$\hiwr$ measurements at intermediate redshift.  Because of these short
comings, we therefore focus on determining which of Churchill's
existing classes best represents the measured properties of each
absorber, in a quantitative sense, instead of running $K$-means
clustering tests from scratch.

To this end, we calculated a matching ``score" that rates how well
each class represents a particular system, with low scores indicating
higher quality matches.  The score for a given class is the sum of the
squared (standardized) deviation between the absorber in question and
zero (the standardized mean, by construction) for each parameter's
distribution---qualitatively similar to a $\chi^2$.

We assigned each system to the class that minimized its match score.
In many cases, we measured only a subset of the nominal five
classification observables.  Since we are calculating a ``best'' match
for each system rather than an absolute match, we simply excluded
those properties from that system's score sum.  In many cases this led to
a classification degeneracy, particularly between classic and
\civnsp-deficient systems where no \civ measurement was available (as was
commonly the case; without \civ these classes are otherwise
indistinguishable, see Churchill's Figure 3).  Where appropriate we
used \civ upper limits to break the degeneracy but in many cases we
could only determine that the system belonged to one of these two
classes.

One complication is that unlike here, \citet{churchill_zoo}
included $\mgiiwr<<0.3\angmath$ systems in their original
standardization procedure.  This reduces the zero point of the
standardized distribution, which in turn increases the renormalized
$y_i$ value for each system above $0.3$\AA.

To compensate for this effect, we re-calculated the standardized means
for each absorption property and each class in the
\citet{churchill_zoo} sample.  Because our literature sample contained
Churchill's data, we could perform this both with and without a
$\mgiiwr<0.3\angmath$ cut applied.  For each parameter we then
measured the offset between means of the cut and full low-redshift
sample.  Then, when classifying each high-redshift absorber we applied
the same offsets in reverse, to capture in a rough sense the effect of
missing systems below $0.3$\AA.

Obviously this crude classification procedure does not account for the
possibility that the classes themselves evolve differently in
redshift, which would manifest as the mean standardized values
changing in redshift.  Our only aim is to provide an objective method
for classifying \mgii absorption systems that is robust to missing
measurements and allows for a first-look study of taxonomy and
evolution of various groupings.

\begin{figure}
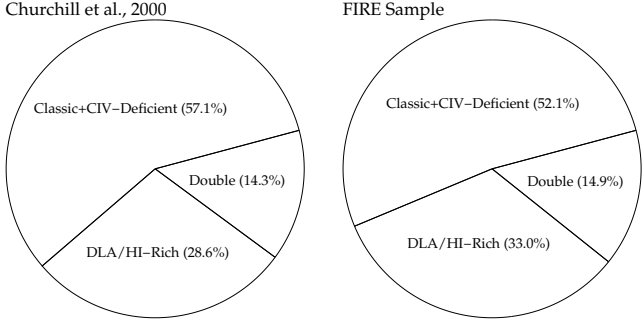

\epsscale{1.20}
\advplotone{Ions}{churchill_pie}{3.7in}{7.0in}
\caption{Percentage classification breakdowns for $\mgiiwr>0.3\angmath$ systems for 
both the low redshift ($0.4<z<1.4$) sample of \citet[][\chclassnall~ systems]{churchill_zoo} and the high redshift 
($z>2$) FIRE sample (\fireclassnall~ systems).  The percentage cuts are remarkably similar given that the universe is in 
vastly different states in the two epochs, separated by $\sim4.8$ Gyr.}
\label{fig:classpie}
\end{figure}

Table \ref{tab:mglist} lists classifications for each system in the
FIRE sample.  If we combine the classic and \civnsp-deficient classes
(because many FIRE systems have no $\civwr$ measurements), we are left
with three classes: classic+\civnsp-deficient, DLA/\hinsp-rich, and double systems.
The fraction of $\mgiiwr\geq0.3\angmath$ systems falling into each of
these three categories is roughly similar at high and low redshift
(Figure \ref{fig:classpie}).  Although the small number of systems
suggests against reading too much into this agreement, the similarity
hints that if these classes result from disparate physical mechanisms,
then the fraction of intersected systems caused by these various
mechanisms has not dramatically evolved over the large redshift range
probed.

We next derived linear densities $dN/dX$ for each absorber class in
isolation.  We did not attempt to adjust the error bars for misclassifications, which surely exist in
non-negligible numbers since many systems have only two or three of
the five absorption properties measured.  We will discuss the ramifications of misclassification in more detail below.

\begin{figure}
\epsscale{1.20}
\advplotone{Ions}{classify_dndx}{5.5in}{4.75in}
\caption{The linear density evolution of
  $0.3\angmath\leq\mgiiwr<1.0\angmath$ systems for the DLA/\hinsp-rich,
  double, and classic+\civnsp-deficient classes of absorbers defined in
  \citet{churchill_zoo}.  Classification depends upon $\mgiiwr$,
  $\hiwr$, $\civwr$, $\feiiwr$ and $\mgiiks$, and is determined using
  the procedure of Section \ref{sec:ions_classify}.  An overwhelming
  majority (\classper\%) of the systems in this $\mgiiwr$ range fall
  into the classic+\civnsp-deficient categories because of relatively
  weak \mgii absorption and low kinematic spreads.  The DLA/\hinsp-rich
  linear density slightly increases with redshift, perhaps a result of
  the rise of the overall DLA population with redshift (Figure
  \ref{fig:dla_dndx}), the increase in $\hiwr/\mgiiwr$ (Figure
  \ref{fig:cdf_ratios}) leading to more DLAs becoming associated
  with $\mgiiwr\lesssim1.0\angmath$ systems, or both.}
\label{fig:classify_dndx}
\end{figure}

Figure \ref{fig:classify_dndx} illustrates $dN/dX$ for
$0.3\angmath\leq\mgiiwr<1.0\angmath$ systems, divided by
classification.  A large majority of systems in this $\mgiiwr$ range
(\classper\%) are classic+\civnsp-deficient.  This is expected since
the classification process considers $\mgiiwr$ and preferentially
assigns strong systems as doubles or DLA\hinsp-rich systems.  But the
classic+\civnsp-deficient set also includes many larger $\mgiiwr$ systems
(including \classnbigcc~ with $\mgiiwr > 0.8\angmath$) that have
$\mgiiwr$ typical of double and DLA\hinsp-rich systems, but were instead classified as
classics on the basis of their small kinematic spreads.  Since the
overall population of absorbers of this strength shows no
statistically significant evidence for evolution from $z\sim0.4$ to
$z\sim5$ (\paperone) and most of these absorbers are
classic+\civnsp-deficient, it is not surprising that the \mgii frequency
for this combined class (bottom panel) also does not significantly
evolve.  Disentangling these two classes to determine their
differential evolution requires more data containing a greater number
of $\civwr$ measurements.

The low incidence of \mgiinsp-weak double systems (\ndoublemid) and high
misclassification probability limit the conclusions we may draw about
their evolution in this range (middle panel).  Likewise the paucity of
DLA systems in this range (\ndlamid) merits caution, although it is
interesting to speculate on the increase in $dN/dX$ towards large
redshift given that both the DLA linear density, most (if not all) of
which appears to be associated with \mgii systems (as discussed later in Section \ref{sec:ions_dlas}), increases over this
redshift range \citep{sdss_dla_dr5}, and the typical $\hiwr$ associated with a given
$\mgiiwr$ increases with redshift (Figure \ref{fig:z_hiratios}).  In
particular, it would be interesting to know whether this increase with
redshift outpaces that of the overall rise, such that a higher
fraction of DLA systems are associated with smaller $\mgiiwr \lesssim
1\angmath$ \mgii systems at high redshift.  Substantially more data
would be required to study this question in detail.

\begin{figure}
\epsscale{1.20}
\advplotone{Ions}{classify_dndx_strong}{6in}{5.25in}
\caption{The linear density evolution of $\mgiiwr\geq1.0\angmath$
  systems for the DLA/\hinsp-rich and double classes of absorbers defined
  in \citet{churchill_zoo}.  The DLA/\hinsp-rich linear density appears
  relatively constant until $z\sim3.5$ before decreasing in the final
  bin.  The double linear density rises by a factor of $3-4$ from
  $z=2$ to 3, and decreases until $z\sim3.5$.  The square points are
  at the new locations of $dN/dX$ if the \classnhuge~
  $\mgiiwr>\hugecut\angmath$ systems are re-classified from
  DLA/\hinsp-rich systems to double systems.  This subset of absorbers,
  for which we do not have $\hiwr$ measurements, possesses both
  unusually large $\feiiwr$
  ($\bar{W}_0^{\lambda2600}$=\classhugefe~\AA) and $\mgiiks$
  ($\bar{\omega}_{2796}$=\classhugeks~\kms).  It is unclear whether
  they belong in the DLA/\hinsp-rich class or double class, or whether
  they constitute an entirely new class of absorber associated with
  physical processes not prevalent at the low $z<1.4$ redshift
  universe studied in \citet{churchill_zoo}.}
\label{fig:classify_dndx_strong}
\end{figure}

Figure \ref{fig:classify_dndx_strong} shows $dN/dX$ for the stronger
$\mgiiwr\geq1.0\angmath$ systems.  Only one classic+\civnsp-deficient
system falls in this range, so we excluded this class from the figure.
Apparently for absorbers with strong $\mgiiwr$ the frequency of DLA/\hinsp-rich 
systems falls from $z=2$ to 5.  The full $\mgiiwr\geq0.3\angmath$ DLA/\hinsp-rich $dN/dX$ remains 
essentially constant over this redshift range.

In contrast, the \mgii frequency of
double systems appears to increase by a factor of $\sim3-4$ from $z=2.2$ to 2.7
before falling until $z\sim3.5$.  Given both empirical evidence
connecting large \mgii absorption to star formation and observations
showing the star formation rate density rising until $z=2-3$ and
falling afterward \citep{bouwens2010,bouwens2011}, it is tempting to
associate double systems with star formation based upon their $dN/dX$
here.  But without $dN/dX$ data for doubles at low $z<2$ redshifts it
is unknown whether the frequency of doubles continues to fall as the
SFR density falls towards $z\rightarrow0$.  Moreover the large error
bars again indicate limitations of our sample size, such that these
evolutionary trends are mostly suggestive and cannot yet be considered
robust.

For example, systems in the lowest redshift bin for these plots
contain no $\hiwr$ or $\civwr$ measurements.  As a result, this bin is
particularly prone to misclassification since $\hiwr$ in particular is
an important diagnostic.  Some systems labeled as DLA/\hinsp-rich in this bin are
therefore marginal classifications based upon extremely strong \feii
absorption, but they also exhibited large kinematic spreads typical of
doubles.  

In fact all but one of the \classnhuge~ $\mgiiwr\geq\hugecut\angmath$ systems
in the FIRE sample (not just those in the lowest redshift bin) were
classified as DLA/\hinsp-rich systems based on their strong \feii absorption
($\bar{W}_0^{\lambda2600}$=\classhugefe~\AA), but we have no $\hiwr$
measurements for any of these systems, and all of them have unusually
large kinematic spreads ($\bar{\omega}_{\lambda2796}$=\classhugeks~\kms).  It
may be that all of these are actually doubles, and the double class as
a whole has evolved between lower redshifts and this epoch.  Figure \ref{fig:ks_plot}, which 
depicts the \mgii and \feii kinematic spreads for all FIRE systems labeled as 
classic+\civnsp-deficient (light gray diamonds), DLA/\hinsp-rich (dark gray circles), and doubles (black triangles), 
provides a case for re-classification:  The large $\mgiiwr$ systems, depicted with 
open circles, occupy a region of $\mgiiks\mbox{-}\omega_{\lambda2600}$ space more heavily 
occupied by double systems. The 
square points on Figure \ref{fig:classify_dndx_strong} represent
$dN/dX$ with the classifications of the $\mgiiwr\geq\hugecut\angmath$
systems changed to double.

\begin{figure}
\epsscale{1.20}
\advplotone{Ions}{ks_plot}{5.5in}{4.25in}
\caption{\mgii and \feii kinematic spreads for the $\mgiiwr\geq0.3\angmath$ FIRE systems.  The light gray diamonds, dark gray circles, and black triangles represent systems classified as classic+\civnsp-deficient, DLA/\hinsp-rich, and double systems, respectively.  The large $\mgiiwr\geq\hugecut\angmath$ systems (enclosed by larger black circles) dominate the upper right portion of the plot.  The matching algorithm predominantly classified these systems as DLA/\hinsp-rich because of strong $\feiiwr$, but they also possess large $\mgiiks$ and occupy a region of $\mgiiks\mbox{-}\omega_{\lambda2600}$ space more heavily occupied by double systems.}
\label{fig:ks_plot}
\end{figure}

\subsection{Connection with DLAs}
\label{sec:ions_dlas}

We showed in Section \ref{sec:ions_hiresults} that $\perdlasfire^{+\perdlashigh}_{-\perdlaslow}$\%
of $\mgiiwr>0.3\angmath$ systems at high redshift
($\bar{z}$=\nhihisubzmean) are associated with DLAs.  We can invert
this question and consider what fraction of high redshift DLAs are
associated with \mgii systems.  \citet{dla_sdss_dr3} provide $dN/dX$
measurements for the general DLA population at redshifts $z=1.7$ to
5.5, which are represented by the gray points in Figure
\ref{fig:dla_dndx}.  The black points represent $dN/dX$ of \mgiinsp-DLAs,
calculated by multiplying total $dN/dX$ for \mgii by the fraction of
\mgii systems exhibiting DLAs in each bin.  For the lowest
\mgiinsp-selected DLA redshift bin, we have no $\nhi$ measurements so we
simply used the fraction for the next highest bin.  This is reasonably
justified since the fraction of DLAs in this $z=2.4609-2.9750$
redshift bin (4/12) is very similar to the fraction for the representative subsample of 
\citet{rt2006} for $z=1-1.5$ (4/13).  The highest \mgiinsp-selected bin
has only one $\nhi$ measurement (a DLA), and is therefore very
uncertain.

\begin{figure}
\epsscale{1.20}
\advplotone{Ions}{dla_dndx}{5.5in}{4.25in}
\caption{The linear densities $dN/dX$ for the general high redshift DLA population \citep[gray points]{dla_sdss_dr3} and the \mgiinsp-selected DLA population (black points).  We calculated the \mgiinsp-selected DLA $dN/dX$ by multiplying the $\mgiiwr>0.3\angmath$ $dN/dX$ for the general \mgii population by the fraction of these systems with $\nhi$ measurements associated with DLAs in each bin.  The first bin had no measurements, and used the fraction from the second.  The largest redshift bin had only one $\nhi$ measurement (a DLA); the errors should be treated with caution.  The plot suggests that an overwhelming majority, if not all, high redshift DLAs have corresponding \mgii absorption.}
\label{fig:dla_dndx}
\end{figure}

This exercise suggests that all (or nearly all) DLAs have accompanying
$\mgiiwr>0.3\angmath$ \mgii absorption.  This is to be expected since
every observed $z>2$ DLA exhibits low-ionization metal line absorption
in rest-frame UV \citep{turnshek89,lu93,wolfe93,lu94}.  Moreover,
\mgii absorption has been found in every high redshift DLA for which
it could have been observed \citep{wolfe05}.  These statements are
also true for low redshift $z<2$ DLAs by construction, since most such DLAs
were selected on the basis of strong \mgii and \feii absorption
\citep{rt2006}.  This result informs our interpretation of Figure
\ref{fig:metals} depicting gas-phase metallicities for the
\mgiinsp-selected (black) and \hinsp-selected (light gray) absorption
systems described in Section \ref{sec:ions_prsub}.  In particular, the
\mgiinsp-selected metallicities for DLAs cannot be inconsistent with
those of the general DLA population if these two groups are largely
the same.

In the representative subsample at low redshifts ($\bar{z}=\rtsubzmean$) from \citet{rt2006} described in Section \ref{sec:ions_rtsub}, a smaller percentage of \mgii systems correspond to
DLAs ($\rtperdlas^{+\rtperdlashigh}_{-\rtperdlaslow}$\%), and these
systems are more commonly associated with Lyman limit systems and/or
sub-DLAs.

It is noteworthy that numerous papers have associated \mgii
systems---particularly the strong variety---with star formation and
outflows (including in our \paperone), yet we find that this population
overlaps very heavily with classical DLAs, which are generally not
thought to result from outflows at all.  Rather DLAs are often taken
as building blocks of present day galaxies \citep{wolfe93}, either as
the early progenitors of galactic disks \citep{prochaska97} or
merging baryonic clumps embedded in dark matter haloes
\citep{haehnelt98,pontzen08}.

The connection between strong \mgii absorption, winds
\citep[e.g.,][]{zibetti2007colors,rubin2010} and DLAs
\citep[e.g.,][]{rt2000} is particularly interesting because low
redshift galaxy-absorber studies see $\mgiiwr>0.3\angmath$ absorption
systems in extended haloes out to $D\sim120h^{-1}$ kpc
\citep{chen2010lowz}, while \mgiinsp-DLAs reside within
$D\lesssim15h^{-1}$ kpc \citep{steidel95} of their respective hosts.
If the strong \mgii systems represent both winds \emph{and} DLAs, then
some fraction of the DLA population would reflect non-gravitational
processes, and also the strong phase of \mgiinsp-absorbing wind evolution would
only fill the halo region nearest to the stellar disk.  This picture may
be incomplete since both $dN/dX$ for the strong absorbers and
empirically derived star formations rates fall from
$z\sim2\rightarrow6$ \citep{bouwens2010,bouwens2011} while the DLA
linear density increases until at least $z\sim5.5$ \citep[see Figure \ref{fig:dla_dndx}, or ][]{dla_sdss_dr3}.  This suggests
that there may be some \mgiinsp-poor DLAs at $z>5$; although such systems
have not been identified, it may be an interesting area for further
study.

One possible alternative is to invoke two populations of strong
absorbers: one corresponding to classical DLAs, and one associated
with star formation driven winds.  This theory is inspired by the
taxonomic classifications of of \citet{churchill_zoo}, and supported
by \citet{bond}, who explore the possibility that strong,
double-troughed \mgii absorbers trace winds.  We see very faint
evidence of evolution in our high redshift $dN/dX$ for the double
systems that is consistent with this interpretation, but cannot be
considered proof on account of the small number statistics.

\subsection{Chemical Evolution}
\label{sec:ions_chemevol}

Figure \ref{fig:cdf_ratios} illustrates how the relative abundance of
\hi at fixed $\mgiiwr$ increases toward higher redshift while the
heavy elements lines remain largely unchanged.  One \emph{might}
interpret this as direct evidence of an increasing metallicity of
\mgii systems toward the present day.  However this picture is
complicated by uncertainties in the degree of saturation in the metal
lines.  At $\mgiiwr=0.3$\AA ~and above one expects some degree of
saturation, particularly for systems with small or unresolved
kinematic spreads.  This effect could in principle mask a decrease in
the metal column densities that tracks the observed change in $\nhi$
from high to low redshifts.

We do estimate lower limits on the abundance directly for systems with
measured \hinsp, finding values consistent with the general DLA
population, and even higher for lower $\nhi$, which correlates
strongly with low $\mgiiwr$.  We also demonstrated that there exists
a large overlap between the DLA and \mgii population, and DLAs evolve
in metallicity as a population, albeit weakly with
a best-fit gradient of $-0.26\pm0.07$ dex per unit redshift and large
scatter \citep{prochaska03}.

The $\hiwr/\mgiiwr$ ratio is similar for weak and strong \mgii systems
(Figure \ref{fig:cdf_ratios_wr}), and $\hiwr/\mgiiwr$ evolves
similarly in redshift for both these sets (Figure
\ref{fig:cdf_ratios_wrz}).  The only discernible difference in metal
line absorption between weak and strong \mgii absorbers is a relative
suppression of high ionization lines (\siivnsp, \civnsp) in the strong
systems.  This may be a straightforward result of ionization effects:
the strong \mgii are more likely to be associated with neutral DLAs,
which are comparatively high in singly ionized species.

The similar $\hiwr/\mgiiwr$ and high metallicities we measure for the
weaker \mgii systems are difficult to reconcile with a scenario where
these systems represent accretion of metal-poor gas from the IGM.
These systems are at least as metal rich as the strong \mgii and
possibly even more so.  However it could follow naturally if the
$0.3\le \mgiiwr\le 1.0$\AA ~absorbers represent the remnants of
previously ejected material, possibly re-accreting as in a galactic
fountain.

In this case the very flat evolution in $dN/dX$ is somewhat surprising
in the absence of fallback, since the cumulative deposition of winds
into the circumgalactic environment should in time increase the \mgii
cross section and hence incidence rate or characteristic abundance.
At $z\sim 5.3$ the Hubble time is just long enough to permit galaxy
formation, wind propagation, and fallback for a few generations.  It
will be interesting to test this at $z\gtrsim 6.2$ as \mgii re-emerges
from the gap between the $H$ and $K$ bands.  As one approaches $z\sim
7$ the timescales for outflow and fallback become challenging, and in
this scenario one would expect the \mgii incidence rate to drop
substantially.

\section{Conclusion}
\label{sec:ions_conclude}

We have presented a large study of chemical abundance properties for
the $z>2$ \mgii systems detected with FIRE in \paperone.  We employ
optical spectra from MagE, MIKE, HIRES and SDSS to measure vacuum
ultraviolet lines such as \hi and \civnsp, as well as singly ionized
states of carbon, silicon, iron, and aluminum.  By combining these
observations with carefully constructed low-redshift control samples,
we perform a longitudinal study of \hi and metals in \mgiinsp-selected
systems from $0<z<5.33$, a period of $>12$ Gyr.  Our main findings are
as follows:

\begin{enumerate}
\item The most significant difference in chemical evolution comes from
  \hinsp, with higher redshift systems associated with much stronger \hi
  column densities.  A K-S test provided only a \nhiks\% probability
  that the low ($z<2$, $\bar{z}$=\rtsubzmean) and high ($z>2$,
  $\bar{z}$=\nhihisubzmean) $\nhi$ samples were drawn from the same
  distribution.  At high redshifts, the fraction of
  $\mgiiwr>0.3\angmath$ systems associated with DLAs
  ($\perdlasfire^{+\perdlashigh}_{-\perdlaslow}$\%) is much larger
  than at lower redshifts
  ($\rtperdlas^{+\rtperdlashigh}_{-\rtperdlaslow}$\%).  All high
  redshift \mgii absorbers are associated with either DLAs or
  sub-DLAs.
\item Comparison between $dN/dX$ for \mgiinsp-selected DLAs and the
  general DLA population at $2<z<5$ shows that a large fraction (if
  not all) of high redshift DLAs have $\mgiiwr>0.4\angmath$
  absorption.  The metallicities for both populations are not
  inconsistent with the hypothesis that the two groups are one and the
  same.
\item \mgii systems associated with sub-DLAs at high redshifts are
  quite metal rich, with some systems possessing lower limits greater
  than one-tenth solar in iron, silicon, and aluminum.
\item Besides \hi and \mginsp, there is no evidence for strong chemical
  evolution in redshift for \mgiinsp-selected systems.  The best
  candidates for moderate chemical evolution are among the high
  ionization states (\siiv 1393, \aliii 1854, and \civ 1548,1550) with
  stronger absorption at lower redshifts (plausibly from less \hi
  shielding), but it is unclear that this evolution is not the result
  of small number counts.
\item Weak $0.3\angmath\leq\mgiiwr\leq1.0\angmath$ and strong
  $\mgiiwr>1.0\angmath$ systems have $\hiwr/\mgiiwr$ ratios that
  are similar in both distribution and redshift evolution.  There is
  some evidence that strong absorbers are associated with weaker high
  ionization states (\siivnsp, \aliiinsp, \civnsp), potentially from shielding
  caused by their higher \hi column densities.
\item Applying the taxonomy defined in \citet{churchill_zoo} to the
  FIRE systems, we find that an overwhelming majority of
  $0.3\angmath\leq\mgiiwr<1.0\angmath$ systems are
  classic+\civnsp-deficient systems (\classper\%).  The linear density of
  this class does not significantly evolve between $2<z<5$.  Strong
  $\mgiiwr\geq1.0\angmath$ systems divide into the DLA/\hinsp-rich and
  double classes.  The strong double $dN/dX$ rises between $z=2$ and 3
  and then falls.  The strong DLA/\hinsp-rich $dN/dX$ falls from $z=2$ to
  5; the full $\mgiiwr\geq0.3\angmath$ DLA/\hinsp-rich $dN/dX$ remains
  essentially constant over this redshift range.
\item The strongest \mgii systems ($\mgiiwr>\hugecut\angmath$;
  \classnhuge~in total) possess unusually strong \feii absorption and
  \mgii kinematic spreads (no $\hiwr$ or $\civwr$ measurements are
  available for these systems).  These systems do not fall nicely into
  any of the five system classes defined in \citet{churchill_zoo}.  It
  is unclear whether they represent DLA/\hinsp-rich systems (as they were
  typically classified), double systems, or an entirely new class generated by
  physical mechanisms not prevalent at $z<1.4$.
\end{enumerate}

The FIRE QSO sample was assembled with the goal of maximizing the
redshift pathlength at higher redshifts $z>3$ in order to provide
better $dN/dX$ estimates in this range.  While the sample accomplished
this stated goal, the high QSO redshifts (typically $z_{QSO}>4$) also
greatly increased the probability that the rest-frame UV and near UV
transitions (e.g., \hi 1215 and \civ1548) of $z\sim2$ systems would
rest blueward of the Lyman break limit of at least one higher redshift
absorber.  As a result, our lowest $\civwr$ and $\hiwr$ measurements
for this $\mgiiwr\geq0.3\angmath$ FIRE sample are $z=2.749$ and
$z=2.593$, respectively.  In addition to targeting high redshift QSOs
to add information on high redshift systems, a new QSO
spectroscopic sample looking to improve upon this study should include
more QSOs with $z_{QSO}\lesssim3$ to better establish the chemical
compositions of $z=2-2.5$ \mgii systems.  It should be possible to use
the SDSS DR7 sample to obtain a list of lower redshift QSOs with
multiple strong \hi systems in this redshift range to strategically
observe QSOs with high probabilities of finding \mgii systems.  If the
\hi distribution of these indicators follow that of the general
population, then this selection process should not bias the chemical
evolution study.

\acknowledgments

We are extremely grateful to the staff of the Magellan Telescopes and
Las Campanas Observatory for their assistance in obtaining the data
presented herein.  This work also benefitted from discussions with
C. Churchill during a brief visit to MIT.  RAS also recognizes the
culturally significant role of the A.J. Burgasser Chair in
Astrophysics.  We gratefully acknowledge financial support from the
NSF under grants AST-0908920 and AST-1109115.  ENS was supported by
the MIT Undergraduate Research Opportunity Program (UROP).

\bibliography{mgii}{}
\bibliographystyle{apj}

\begin{deluxetable}{l c c c c c}
\tablecaption{ ~ \\ [-2mm] Complimentary Optical Spectra}
\tablehead{ \colhead{Object} & \colhead{$z_{QSO}$} & \colhead{Instrument} & \colhead{$t_{exp}$} & \colhead{$\lambda$ Range\tablenotemark{a}} & \colhead{SNR\tablenotemark{b}} \\ \colhead{} & \colhead{} & \colhead{} & \colhead{(s)} & \colhead{(\AA)} & \colhead{pix$^{-1}$} }
\startdata
Q0000$-$26 & 4.10 & MagE & 900 & 6197$-$10283 & 23.0 \\
BR0004$-$6224 & 4.51 & (none) & ..... & ..... & ..... \\
BR0016$-$3544 & 4.15 & (none) & ..... & ..... & ..... \\
SDSS0106+0048 & 4.45 & SDSS & 2700 & 6624$-$9219 & 11.0 \\
SDSS0113$-$0935 & 3.67 & SDSS & 3601 & 5674$-$9221 & 14.8 \\
SDSS0127$-$0045 & 4.08 & SDSS & 5706 & 6178$-$9221 & 16.9 \\
SDSS0140$-$0839 & 3.71 & MagE & 900 & 5729$-$10283 & 38.5 \\
SDSS0203+0012 & 5.85 & (none) & ..... & ..... & ..... \\
\enddata
\tablenotetext{a}{Minimum wavelength listed is Ly-$\alpha$ emission wavelength of QSO.}
\tablenotetext{b}{Median signal-to-noise ratio redward of Ly-$\alpha$ forest.}
\label{tab:optical}
\end{deluxetable}

\begin{deluxetable*}{ c l c c c c c }
\tablewidth{0pc}
\tablecaption{ ~ \\ [-2mm] Absorption Properties for the FIRE \mgiinsp-Selected Sample: Mg Ions}
\tablehead{ \colhead{Index \#} & \colhead{Sightline} & \colhead{$z$} & $W_0^{\lambda2796}$ & $W_0^{\lambda2803}$ & $W_0^{\lambda2852}$\tablenotemark{a} & \colhead{Class} \\ \colhead{} & \colhead{} & \colhead{}& \colhead{(\AA)} & \colhead{(\AA)} & \colhead{(\AA)} & \colhead{} }\startdata

1\tablenotemark{b} & Q0000$-$26 & 2.184 & $0.162\pm0.031$ & $0.112\pm0.026$ & $<$ 0.114~ & ....... \\
2~~ & Q0000$-$26 & 3.390 & $1.356\pm0.016$ & $1.145\pm0.016$ & $0.060\pm0.014$ & DLA/\hi-Rich \\
3~~ & BR0004$-$6224 & 2.663 & $0.260\pm0.045$ & $0.140\pm0.036$ & $<$ 0.079~ & ....... \\
4~~ & BR0004$-$6224 & 2.908 & $0.596\pm0.047$ & $0.183\pm0.028$ & $<$ 0.133~ & Classic+\civnsp-Deficient \\
5~~ & BR0004$-$6224 & 2.959 & $0.569\pm0.063$ & $0.669\pm0.047$ & $0.161\pm0.043$ & Classic+\civnsp-Deficient \\
6~~ & BR0004$-$6224 & 3.203 & $0.558\pm0.029$ & $0.548\pm0.026$ & $<$ 0.081~ & Classic+\civnsp-Deficient \\
7~~ & BR0004$-$6224 & 3.694 & $0.236\pm0.042$ & $0.234\pm0.019$ & $<$ 1.030\tablenotemark{c} & ....... \\
8~~ & BR0004$-$6224 & 3.776 & $1.045\pm0.046$ & $1.049\pm0.043$ & ... & DLA/\hi-Rich \\
\enddata
\label{tab:mglist}
\end{deluxetable*}

\setcounter{table}{1}
\begin{deluxetable*}{ c l c c c c c }
\tablewidth{0pc}
\tablecaption{ ~ \\ [-2mm] Absorption Properties for the FIRE \mgiinsp-Selected Sample: Mg Ions (\emph{Continued})}
\tablehead{ \colhead{Index \#} & \colhead{Sightline} & \colhead{$z$} & $W_0^{\lambda2796}$ & $W_0^{\lambda2803}$ & $W_0^{\lambda2852}$\tablenotemark{a} & \colhead{Class} \\ \colhead{} & \colhead{} & \colhead{}& \colhead{(\AA)} & \colhead{(\AA)} & \colhead{(\AA)} & \colhead{} }\startdata

61~~ & SDSS1110+0244 & 2.223 & $0.205\pm0.024$ & $0.121\pm0.029$ & $<$ 0.090~ & ....... \\
62~~ & SDSS1305+0521 & 2.302 & $1.993\pm0.095$ & $1.533\pm0.095$ & $<$ 0.321~ & DLA/\hi-Rich \\
63~~ & SDSS1305+0521 & 2.753 & $0.378\pm0.040$ & $0.319\pm0.037$ & $<$ 0.115~ & Classic+\civnsp-Deficient \\
64~~ & SDSS1305+0521 & 3.235 & $0.328\pm0.025$ & $0.130\pm0.025$ & $<$ 0.118~ & \civnsp-Deficient \\
65~~ & SDSS1305+0521 & 3.680 & $1.781\pm0.068$ & $1.583\pm0.039$ & ... & DLA/\hi-Rich \\
66~~ & SDSS1306+0356 & 2.533 & $3.307\pm0.101$ & $3.019\pm0.088$ & $<$ 0.126~ & DLA/\hi-Rich \\
67~~ & SDSS1306+0356 & 3.490 & $0.648\pm0.031$ & $0.526\pm0.044$ & $<$ 0.229~ & Classic+\civnsp-Deficient \\
68~~ & SDSS1306+0356 & 4.615 & $0.983\pm0.078$ & $0.724\pm0.038$ & $0.076\pm0.021$ & DLA/\hi-Rich \\
\enddata
\tablenotetext{a}{Upper limits are 3$\sigma$.}
\tablenotetext{b}{Suspected false positive.}
\tablenotetext{c}{Blended line.}
\tablenotetext{d}{$\mgiiwr$ and $W_0^{\lambda2803}$ values adjusted relative to Paper I.}
\nonumber
\end{deluxetable*}

\begin{deluxetable}{c l c c c c c c c c}
\tablecaption{ ~ \\ [-2mm] The Low Redshift Compilation Sample\tablenotemark{a}}
\tablehead{ \colhead{ID \#} & \colhead{Sightline} & \colhead{$z_{QSO}$} & \colhead{$z$} & \colhead{$W_0^{\lambda2796}$} & \colhead{$W_0^{\lambda1215}$\tablenotemark{b}} & \colhead{$W_0^{\lambda1393}$\tablenotemark{b}} & \colhead{$W_0^{\lambda1548}$\tablenotemark{b}} & \colhead{$W_0^{\lambda2600}$\tablenotemark{b}}  & \colhead{Ref.\tablenotemark{c}} \\  \colhead{} & \colhead{} & \colhead{} & \colhead{} & \colhead{(\AA)} & \colhead{(\AA)} & \colhead{(\AA)} & \colhead{(\AA)} & \colhead{(\AA)} & \colhead{} }
\startdata
L1 & 0001+087 & 3.243 & 1.085 & $0.249\pm0.034$ & ..... & ..... & ..... & $<0.264$~~ & 1 \\
L2 & 0001+087 & 3.243 & 1.416 & $0.584\pm0.046$ & ..... & ..... & ..... & $<0.228$~~ & 1 \\
L3 & 0002+051 & 1.899 & 0.591 & $0.103\pm0.000$ & ..... & ..... & ..... & $<0.012$~~ & 2,3 \\
L4 & 0002+051 & 1.900 & 0.851 & $1.043\pm0.000$ & $2.470\pm0.080$ & $<0.918$~~ & $1.260\pm0.060$ & $0.419\pm0.022$ & 2,4,5 \\
L5 & 0002+051 & 1.900 & 0.956 & $0.052\pm0.000$ & $0.821\pm0.003$ & $<0.869$~~ & $0.509\pm0.001$ & $<0.005$~~ & 2,3,4 \\
L6 & 0002$-$422 & 2.758 & 0.837 & $4.683\pm0.381$ & ..... & ..... & ..... & $2.940\pm0.272$ & 6,7 \\
L7 & 0002$-$422 & 2.758 & 1.541 & $0.480\pm0.043$ & ..... & ..... & $0.710\pm0.000$ & $<0.275$~~ & 7 \\
L8 & 0004+171 & 2.890 & 0.807 & $1.533\pm0.061$ & ..... & ..... & ..... & ..... & 1 \\
\enddata
\tablenotetext{a}{This table is available in a machine-readable form in the online journal. A portion is shown here for guidance regarding its form and content.}
\tablenotetext{b}{Upper limits are 5$\sigma$.}
\tablenotetext{c}{REFERENCES.--(1) \citet{sargent89}; (2) \citet{churchill_2000a}; (3) \citet{churchill_weakmgii}; (4) \citet{hst_mgii3}; (5) \citet{steidelandsargent}; (6) \citet{sargent1979}; (7) \citet{lanzetta_mgii}; (8) \citet{sargentmgii}; (9) \citet{barthel90}; (10) \citet{steidel90}; (11) \citet{sl1996}; (12) \citet{sargent82a}; (13) \citet{aldcroft94}; (14) \citet{ysb1982b}; (15) \citet{petitjean94}; (16) \citet{wright1982}; (17) \citet{petitjean1990}; (18) \citet{ysb1982a}; (19) \citet{tytler_mgii}; (20) \citet{Foltz1986}; (21) \citet{RS1983}; (22) \citet{hst_mgii2}; (23) \citet{hst_mgii1}.
}
\tablenotetext{d}{Blended line.}
\tablenotetext{e}{FeII 2382.}
\tablenotetext{f}{FeII 2586.}
\tablenotetext{g}{FeII 1608.}
\label{tab:lowzsample}
\end{deluxetable}

\begin{deluxetable}{c l c c c c c c c}
\tablecaption{ ~ \\ [-2mm] $\nhi$ and Metallicity Measurements, FIRE sample}
\tablehead{ \colhead{Index \#} & \colhead{Sightline} & \colhead{$z$} & \colhead{$\lognhi$} & \colhead{[Fe/H]\tablenotemark{a}} & \colhead{[C/H]\tablenotemark{b}} & \colhead{[Si/H]\tablenotemark{c}} & \colhead{[Al/H]\tablenotemark{d}} & \colhead{[Mg/H]\tablenotemark{e}} \\  \colhead{} & \colhead{} & \colhead{} & \colhead{(log \pcmsq)} & \colhead{} & \colhead{} & \colhead{} & \colhead{} & \colhead{} }
\startdata
2~ & Q0000$-$26 & 3.390 & $21.40\pm0.05$ & $>-2.524$ & ......... & $>-1.330$ & $>-2.920$ & $>-3.045$ \\
13~ & SDSS0106+0048 & 3.729 & $19.45\pm0.15$ & $>-1.442$ & ......... & $>-0.858$ & $>-0.855$ & $>-1.379$ \\
15~ & SDSS0113$-$0935 & 3.544 & $19.05\pm0.10$ & ......... & ......... & ......... & ......... & $>-1.692$ \\
16~ & SDSS0113$-$0935 & 3.617 & $19.40\pm0.15$ & ......... & $>-1.497$ & $>-1.523$ & ......... & $>-1.731$ \\
18~ & SDSS0127$-$0045 & 2.945 & $20.03\pm0.15$ & $>-2.052$ & ......... & ......... & $>-1.228$ & $>-1.614$ \\
19~ & SDSS0127$-$0045 & 3.168 & $19.20\pm0.15$ & ......... & ......... & ......... & ......... & $>-1.889$ \\
20~ & SDSS0127$-$0045 & 3.728 & $21.20\pm0.10$ & $>-2.845$ & $>-3.144$ & $>-2.708$ & $>-2.890$ & $>-3.073$ \\
22~ & SDSS0140$-$0839 & 3.081 & $19.25\pm0.20$ & $-1.046\pm0.312$ & ......... & $>-0.952$ & $>-0.968$ & $>-1.453$ \\
\enddata
\tablenotetext{a}{\feiinsp.}
\tablenotetext{b}{\ciinsp.}
\tablenotetext{c}{\siiinsp.}
\tablenotetext{d}{\aliinsp.}
\tablenotetext{e}{\mgiinsp.}
\tablenotetext{f}{[O/H] = $-2.037\pm0.247$ (\oinsp).}
\label{tab:nhi_table}
\end{deluxetable}

\clearpage
\begin{deluxetable*}{ c l c c c}
\tablewidth{0pc}
\tablecaption{ ~ \\ [-2mm] Absorption Properties for the FIRE \mgiinsp-Selected Sample: \hi 1215 and \oi 1302}
\tablehead{ \colhead{Index \#} & \colhead{Sightline} & \colhead{$z$} & $W_0^{\lambda1215}$ & $W_0^{\lambda1302}$\tablenotemark{a} \\ \colhead{} & \colhead{} & \colhead{}& \colhead{(\AA)} & \colhead{(\AA)} }\startdata

2~~ & Q0000$-$26 & 3.390 & $33.182\pm0.093$ & ... \\
13~~ & SDSS0106+0048 & 3.729 & $3.863\pm0.139$ & ... \\
15~~ & SDSS0113$-$0935 & 3.544 & $2.470\pm0.068$ & $<$ 0.142~ \\
16~~ & SDSS0113$-$0935 & 3.617 & $3.651\pm0.061$ & $<$ 0.156~ \\
18~~ & SDSS0127$-$0045 & 2.945 & $8.102\pm0.176$ & ... \\
19~~ & SDSS0127$-$0045 & 3.168 & $2.916\pm0.088$ & ... \\
20~~ & SDSS0127$-$0045 & 3.728 & $26.937\pm0.357$ & ... \\
22~~ & SDSS0140$-$0839 & 3.081 & $3.086\pm0.026$ & ... \\
\enddata
\tablenotetext{a}{Upper limits are 3$\sigma$.}
\tablenotetext{b}{Blended line.}
\label{tab:olist}
\end{deluxetable*}

\begin{deluxetable*}{ c l c c c c c c c}
\tablewidth{0pc}
\tablecaption{ ~ \\ [-2mm] Absorption Properties for the FIRE \mgiinsp-Selected Sample: Fe Ions}
\tablehead{ \colhead{Index \#} & \colhead{Sightline} & \colhead{$z$} & $W_0^{\lambda1608}$\tablenotemark{a} & $W_0^{\lambda2344}$\tablenotemark{a} & $W_0^{\lambda2374}$\tablenotemark{a} & $W_0^{\lambda2382}$\tablenotemark{a} & $W_0^{\lambda2586}$\tablenotemark{a} & $W_0^{\lambda2600}$\tablenotemark{a} \\ \colhead{} & \colhead{} & \colhead{}& \colhead{(\AA)} & \colhead{(\AA)} & \colhead{(\AA)} & \colhead{(\AA)} & \colhead{(\AA)} & \colhead{(\AA)} }\startdata

1\tablenotemark{b} & Q0000$-$26 & 2.184 & ... & $<$ 0.056~ & $<$ 0.056~ & $<$ 0.052~ & $<$ 0.471\tablenotemark{c} & $<$ 0.106~ \\
2~~ & Q0000$-$26 & 3.390 & $0.190\pm0.010$ & $0.313\pm0.031$ & $0.302\pm0.021$ & $0.514\pm0.017$ & $0.329\pm0.025$ & $0.523\pm0.016$ \\
3~~ & BR0004$-$6224 & 2.663 & ... & $<$ 0.236~ & $<$ 0.266~ & $<$ 0.334~ & $<$ 0.258~ & $<$ 0.264~ \\
4~~ & BR0004$-$6224 & 2.908 & ... & $<$ 0.241~ & $<$ 0.277\tablenotemark{c} & $<$ 0.199~ & $<$ 0.118~ & $<$ 0.131~ \\
5~~ & BR0004$-$6224 & 2.959 & ... & $0.318\pm0.058$ & $<$ 0.203~ & $0.549\pm0.084$ & $<$ 0.293\tablenotemark{c} & $0.557\pm0.047$ \\
6~~ & BR0004$-$6224 & 3.203 & ... & $0.284\pm0.054$ & $<$ 0.099~ & $0.463\pm0.049$ & $0.109\pm0.031$ & $<$ 0.588\tablenotemark{c} \\
7~~ & BR0004$-$6224 & 3.694 & ... & $<$ 0.081~ & $<$ 0.180\tablenotemark{c} & $<$ 0.335~ & $<$ 0.133~ & $<$ 0.090~ \\
8~~ & BR0004$-$6224 & 3.776 & ... & $0.683\pm0.044$ & $0.209\pm0.031$ & $0.730\pm0.031$ & $0.446\pm0.044$ & $0.697\pm0.032$ \\
\enddata
\label{tab:felist}
\end{deluxetable*}

\setcounter{table}{5}
\begin{deluxetable*}{ c l c c c c c c c}
\tablewidth{0pc}
\tablecaption{ ~ \\ [-2mm] Absorption Properties for the FIRE \mgiinsp-Selected Sample: Fe Ions (\emph{Continued})}
\tablehead{ \colhead{Index \#} & \colhead{Sightline} & \colhead{$z$} & $W_0^{\lambda1608}$\tablenotemark{a} & $W_0^{\lambda2344}$\tablenotemark{a} & $W_0^{\lambda2374}$\tablenotemark{a} & $W_0^{\lambda2382}$\tablenotemark{a} & $W_0^{\lambda2586}$\tablenotemark{a} & $W_0^{\lambda2600}$\tablenotemark{a} \\ \colhead{} & \colhead{} & \colhead{}& \colhead{(\AA)} & \colhead{(\AA)} & \colhead{(\AA)} & \colhead{(\AA)} & \colhead{(\AA)} & \colhead{(\AA)} }\startdata

61~~ & SDSS1110+0244 & 2.223 & ... & $<$ 0.061~ & ... & ... & $<$ 0.095~ & $<$ 0.084~ \\
62~~ & SDSS1305+0521 & 2.302 & ... & $1.112\pm0.109$ & $<$ 0.389~ & $1.161\pm0.147$ & $0.480\pm0.160$ & $1.407\pm0.099$ \\
63~~ & SDSS1305+0521 & 2.753 & ... & $<$ 0.311~ & $<$ 0.123~ & $<$ 0.116~ & $<$ 0.093~ & $<$ 0.141~ \\
64~~ & SDSS1305+0521 & 3.235 & $<$ 0.181~ & $0.146\pm0.040$ & $<$ 0.114~ & $0.112\pm0.037$ & $<$ 0.101~ & $<$ 0.141~ \\
65~~ & SDSS1305+0521 & 3.680 & $0.443\pm0.129$ & $0.891\pm0.048$ & $0.367\pm0.059$ & $0.932\pm0.050$ & $0.691\pm0.037$ & $0.984\pm0.042$ \\
66~~ & SDSS1306+0356 & 2.533 & ... & ... & ... & ... & $0.910\pm0.087$ & $2.214\pm0.088$ \\
67~~ & SDSS1306+0356 & 3.490 & ... & $<$ 0.104~ & $<$ 0.110~ & $0.154\pm0.038$ & $<$ 0.139~ & $<$ 0.101~ \\
68~~ & SDSS1306+0356 & 4.615 & ... & $0.227\pm0.028$ & $<$ 0.097~ & $0.332\pm0.029$ & ... & ... \\
\enddata
\tablenotetext{a}{Upper limits are 3$\sigma$.}
\tablenotetext{b}{Suspected false positive.}
\tablenotetext{c}{Blended line.}
\nonumber
\end{deluxetable*}

\clearpage
\begin{deluxetable*}{ c l c c c c c c c}
\tablewidth{0pc}
\tablecaption{ ~ \\ [-2mm] Absorption Properties for the FIRE \mgiinsp-Selected Sample: Si Ions}
\tablehead{ \colhead{Index \#} & \colhead{Sightline} & \colhead{$z$} & $W_0^{\lambda1260}$ & $W_0^{\lambda1304}$\tablenotemark{a} & $W_0^{\lambda1393}$\tablenotemark{a} & $W_0^{\lambda1402}$\tablenotemark{a} & $W_0^{\lambda1526}$\tablenotemark{a} & $W_0^{\lambda1808}$\tablenotemark{a} \\ \colhead{} & \colhead{} & \colhead{}& \colhead{(\AA)} & \colhead{(\AA)} & \colhead{(\AA)} & \colhead{(\AA)} & \colhead{(\AA)} & \colhead{(\AA)} }\startdata

2~~ & Q0000$-$26 & 3.390 & ... & ... & ... & ... & $0.340\pm0.011$ & $0.198\pm0.012$ \\
7~~ & BR0004$-$6224 & 3.694 & ... & ... & ... & ... & ... & $<$ 0.176~ \\
8~~ & BR0004$-$6224 & 3.776 & ... & ... & ... & ... & ... & $<$ 0.162~ \\
12~~ & BR0016$-$3544 & 3.757 & ... & ... & ... & ... & ... & $<$ 0.079~ \\
13~~ & SDSS0106+0048 & 3.729 & ... & ... & ... & ... & $0.305\pm0.060$ & $<$ 0.080~ \\
14\tablenotemark{b} & SDSS0113$-$0935 & 2.825 & ... & ... & ... & ... & $<$ 0.173~ & $<$ 0.161~ \\
15~~ & SDSS0113$-$0935 & 3.544 & $<$ 0.300\tablenotemark{c} & $<$ 0.136~ & $0.138\pm0.045$ & $0.216\pm0.036$ & $<$ 0.134~ & $<$ 0.204~ \\
16~~ & SDSS0113$-$0935 & 3.617 & $0.313\pm0.050$ & $<$ 0.164~ & $0.480\pm0.040$ & $0.580\pm0.044$ & $<$ 0.200\tablenotemark{c} & $<$ 0.112~ \\
\enddata
\tablenotetext{a}{Upper limits are 3$\sigma$.}
\tablenotetext{b}{Suspected false positive.}
\tablenotetext{c}{Blended line.}
\label{tab:silist}
\end{deluxetable*}

\begin{deluxetable*}{ c l c c c c c c c}
\tablewidth{0pc}
\tablecaption{ ~ \\ [-2mm] Absorption Properties for the FIRE \mgiinsp-Selected Sample: C and Al Ions}
\tablehead{ \colhead{Index \#} & \colhead{Sightline} & \colhead{$z$} & $W_0^{\lambda1334}$\tablenotemark{a} & $W_0^{\lambda1548}$\tablenotemark{a} & $W_0^{\lambda1550}$\tablenotemark{a} & $W_0^{\lambda1670}$\tablenotemark{a} & $W_0^{\lambda1854}$\tablenotemark{a} & $W_0^{\lambda1862}$\tablenotemark{a} \\ \colhead{} & \colhead{} & \colhead{}& \colhead{(\AA)} & \colhead{(\AA)} & \colhead{(\AA)} & \colhead{(\AA)} & \colhead{(\AA)} & \colhead{(\AA)} }\startdata

2~~ & Q0000$-$26 & 3.390 & ... & $0.959\pm0.012$ & $0.797\pm0.012$ & $0.290\pm0.018$ & $0.087\pm0.023$ & $0.101\pm0.025$ \\
7~~ & BR0004$-$6224 & 3.694 & ... & ... & ... & ... & $<$ 0.205~ & $<$ 0.238~ \\
8~~ & BR0004$-$6224 & 3.776 & ... & ... & ... & ... & $<$ 0.208~ & $<$ 0.201~ \\
12~~ & BR0016$-$3544 & 3.757 & ... & ... & ... & ... & $<$ 0.111~ & $<$ 0.096~ \\
13~~ & SDSS0106+0048 & 3.729 & ... & $0.989\pm0.106$ & $0.879\pm0.084$ & $0.464\pm0.097$ & $0.089\pm0.028$ & $<$ 0.080~ \\
14\tablenotemark{b} & SDSS0113$-$0935 & 2.825 & ... & $<$ 0.161~ & $<$ 0.165~ & $<$ 0.170~ & $<$ 0.143~ & $<$ 0.152~ \\
15~~ & SDSS0113$-$0935 & 3.544 & $<$ 0.123~ & $0.288\pm0.036$ & $0.203\pm0.043$ & $<$ 0.181~ & $<$ 0.063~ & $<$ 0.079~ \\
16~~ & SDSS0113$-$0935 & 3.617 & $0.388\pm0.055$ & $0.704\pm0.042$ & $0.390\pm0.037$ & $<$ 0.370~ & $<$ 0.095~ & $<$ 0.075~ \\
\enddata
\tablenotetext{a}{Upper limits are 3$\sigma$.}
\tablenotetext{b}{Suspected false positive.}
\tablenotetext{c}{Blended line.}
\label{tab:clist}
\end{deluxetable*}

\clearpage
\begin{deluxetable*}{c l c c c c c c c}
\tablecaption{ ~ \\ [-2mm] Kinematic Properties, FIRE sample}
\tablehead{ \colhead{Index} & \colhead{Sightline} & \colhead{$z$}  & \colhead{$\Delta v_{\lambda2796}$} & \colhead{$\omega_{\lambda2796}$}  & \colhead{$\Delta v_{\lambda2600}$} & \colhead{$\omega_{\lambda2600}$}  & \colhead{$\Delta v_{\lambda1548}$} & \colhead{$\omega_{\lambda1548}$}  \\ [1mm]   \colhead{} & \colhead{} & \colhead{}  & \colhead{(\kms)} & \colhead{(\kms)} & \colhead{(\kms)} & \colhead{(\kms)} & \colhead{(\kms)} & \colhead{(\kms)}}
\startdata
1\tablenotemark{a} & Q0000$-$26 & 2.184 & $175.2\pm17.5$ & $61.2\pm8.6$ & ..... & ..... & ..... & ..... \\ 
2~~ & Q0000$-$26 & 3.390 & $262.7\pm26.3$ & $52.2\pm1.1$ & $150.1\pm15.0$ & $36.0\pm1.0$ & $290.1\pm29.0$ & $78.7\pm1.2$ \\ 
3~~ & BR0004$-$6224 & 2.663 & $62.6\pm8.8$ & $34.1\pm5.6$ & ..... & ..... & ..... & ..... \\ 
4~~ & BR0004$-$6224 & 2.908 & $112.6\pm11.3$ & $40.1\pm3.7$ & ..... & ..... & ..... & ..... \\ 
5~~ & BR0004$-$6224 & 2.959 & $125.1\pm12.5$ & $26.4\pm3.7$ & $150.1\pm15.0$ & $36.7\pm2.7$ & ..... & ..... \\ 
6~~ & BR0004$-$6224 & 3.203 & $100.1\pm10.0$ & $29.1\pm1.6$ & $100.1\pm10.0$\tablenotemark{b} & $29.8\pm4.2$\tablenotemark{b} & ..... & ..... \\ 
7~~ & BR0004$-$6224 & 3.694 & $100.1\pm10.0$ & $32.7\pm4.8$ & ..... & ..... & ..... & ..... \\ 
8~~ & BR0004$-$6224 & 3.776 & $175.2\pm17.5$ & $43.8\pm2.5$\tablenotemark{c} & $125.1\pm12.5$ & $32.5\pm1.5$ & ..... & ..... \\
\enddata
\label{tab:ks}
\end{deluxetable*}

\setcounter{table}{8}
\begin{deluxetable*}{c l c c c c c c c}
\tablecaption{ ~ \\ [-2mm] Kinematic Properties, FIRE sample (\emph{Continued})}
\tablehead{ \colhead{Index} & \colhead{Sightline} & \colhead{$z$}  & \colhead{$\Delta v_{\lambda2796}$} & \colhead{$\omega_{\lambda2796}$}  & \colhead{$\Delta v_{\lambda2600}$} & \colhead{$\omega_{\lambda2600}$}  & \colhead{$\Delta v_{\lambda1548}$} & \colhead{$\omega_{\lambda1548}$}  \\ [1mm]   \colhead{} & \colhead{} & \colhead{}  & \colhead{(\kms)} & \colhead{(\kms)} & \colhead{(\kms)} & \colhead{(\kms)} & \colhead{(\kms)} & \colhead{(\kms)}}
\startdata
61~~ & SDSS1110+0244 & 2.223 & $137.6\pm13.8$ & $36.7\pm3.1$ & ..... & ..... & ..... & ..... \\ 
62~~ & SDSS1305+0521 & 2.302 & $237.7\pm23.8$ & ..... & $200.2\pm20.0$ & ..... & ..... & ..... \\ 
63~~ & SDSS1305+0521 & 2.753 & $162.6\pm16.3$ & $41.1\pm3.3$ & ..... & ..... & ..... & ..... \\ 
64~~ & SDSS1305+0521 & 3.235 & $125.1\pm12.5$ & $34.6\pm2.0$ & $150.2\pm15.0$\tablenotemark{d} & $37.3\pm9.5$\tablenotemark{d} & ..... & ..... \\ 
65~~ & SDSS1305+0521 & 3.680 & $387.9\pm38.8$ & $83.3\pm12.2$ & $175.1\pm17.5$ & $42.6\pm14.1$ & ..... & ..... \\ 
66~~ & SDSS1306+0356 & 2.533 & $587.9\pm58.8$ & ..... & $562.9\pm56.3$ & $117.9\pm15.4$ & ..... & ..... \\ 
67~~ & SDSS1306+0356 & 3.490 & $175.1\pm17.5$ & $45.8\pm1.8$ & $175.1\pm17.5$\tablenotemark{d} & $51.3\pm6.4$\tablenotemark{d} & ..... & ..... \\ 
68~~ & SDSS1306+0356 & 4.615 & $225.2\pm22.5$ & $45.4\pm7.9$ & $137.6\pm13.8$\tablenotemark{d} & $28.0\pm1.9$\tablenotemark{d} & $300.2\pm30.0$ & $59.9\pm1.8$ \\
\enddata
\tablenotetext{a}{Suspected false positive.}
\tablenotetext{b}{FeII 2586.}
\tablenotetext{c}{MgII 2803.}
\tablenotetext{d}{FeII 2382.}
\tablenotetext{e}{FeII 2374.}
\tablenotetext{f}{FeII 2344.}
\end{deluxetable*}

\begin{deluxetable}{c c c c c c}
\tablecaption{ ~ \\ [-2mm] Comparison Samples Statistics: Two Redshift Bins}
\tablehead{ \colhead{Ion} & \colhead{$\bar{z}$} & \colhead{Detections} & \colhead{Upper Limits\tablenotemark{a}\tablenotemark{b}} & \colhead{Min $W_{\hbox{Ion}}$} & \colhead{Max $W_{\hbox{Ion}}$} \\   \colhead{} & \colhead{} & \colhead{} & \colhead{} & \colhead{(\AA)} & \colhead{(\AA)} }
\startdata
 ~ \\ [-2mm] \multicolumn{6}{c}{Low Redshift, $0<z<2$} \\ [1mm]
\hline
 ~ \\ [-2mm] HI 1215 & 0.864 & 18 & 2~(1) & $1.09\pm0.000$ & $11.95\pm0.26$ \\
MgII 2796 & 1.128 & 272 & 0 & $0.30\pm0.04$ & $5.80\pm0.08$ \\
MgI 2852 & 1.128 & 46 & 214~(5) & $<0.05$ & $1.36\pm0.11$ \\
FeII 2374 & 1.215 & 36 & 137~(5) & $<0.07$ & $2.14\pm0.08$ \\
FeII 2586 & 1.190 & 66 & 139~(3) & $<0.03$ & $2.60\pm0.14$ \\
\enddata
\tablenotetext{a}{Upper Limits are 5$\sigma$.}
\tablenotetext{b}{Values in parentheses are upper limits derived from blended lines.}
\label{tab:sample_counts}
\end{deluxetable}

\begin{deluxetable}{c c c c c c}
\tablecaption{ ~ \\ [-2mm] Comparison Samples Statistics: Three Redshift Bins}
\tablehead{ \colhead{Ion} & \colhead{$\bar{z}$} & \colhead{Detections} & \colhead{Upper Limits\tablenotemark{a}\tablenotemark{b}} & \colhead{Min $W_{\hbox{Ion}}$} & \colhead{Max $W_{\hbox{Ion}}$} \\   \colhead{} & \colhead{} & \colhead{} & \colhead{} & \colhead{(\AA)} & \colhead{(\AA)} }
\startdata
 ~ \\ [-2mm] \multicolumn{6}{c}{Low Redshift, $0<z<1.5$} \\ [1mm]
\hline
 ~ \\ [-2mm] HI 1215 & 0.752 & 17 & 1 & $1.09\pm0.000$ & $11.15\pm1.09$ \\
MgII 2796 & 0.968 & 217 & 0 & $0.30\pm0.04$ & $4.68\pm0.38$ \\
MgI 2852 & 0.973 & 37 & 172~(3) & $<0.05$ & $1.36\pm0.11$ \\
FeII 2374 & 1.008 & 26 & 101~(2) & $<0.07$ & $1.09\pm0.00$ \\
FeII 2586 & 0.997 & 47 & 107~(3) & $<0.03$ & $1.90\pm0.02$ \\
\enddata
\tablenotetext{a}{Upper Limits are 5$\sigma$.}
\tablenotetext{b}{Values in parentheses are upper limits derived from blended lines.}
\label{tab:sample_counts3}
\end{deluxetable}

\begin{deluxetable}{c c c c c c}
\tablecaption{ ~ \\ [-2mm] Weak and Strong $W_0^{\lambda2796}$ Comparison Samples Statistics}
\tablehead{ \colhead{Ion} & \colhead{$\bar{z}$} & \colhead{Detections} & \colhead{Upper Limits\tablenotemark{a}\tablenotemark{b}} & \colhead{Min $W_{\hbox{Ion}}$} & \colhead{Max $W_{\hbox{Ion}}$} \\   \colhead{} & \colhead{} & \colhead{} & \colhead{} & \colhead{(\AA)} & \colhead{(\AA)} }
\startdata
 ~ \\ [-2mm] \multicolumn{6}{c}{$0.3\angmath\leq W_0^{\lambda2796}\leq1.0\angmath$} \\ [1mm]
\hline
 ~ \\ [-2mm] HI 1215 & 2.805 & 21 & 0 & $1.09\pm0.000$ & $26.94\pm0.36$ \\
MgII 2796 & 2.129 & 119 & 0 & $0.30\pm0.04$ & $0.99\pm0.00$ \\
MgI 2852 & 2.116 & 8 & 108~(2) & $0.02\pm0.004$ & $0.32\pm0.04$ \\
FeII 2374 & 2.283 & 11 & 77~(3) & $<0.03$ & $0.59\pm0.05$ \\
FeII 2586 & 2.204 & 24 & 70~(1) & $<0.06$ & $0.75\pm0.04$ \\
\enddata
\tablenotetext{a}{Upper Limits are 5$\sigma$.}
\tablenotetext{b}{Values in parentheses are upper limits derived from blended lines.}
\label{tab:sample_counts_wr}
\end{deluxetable}

\begin{deluxetable}{c c c c c c}
\tablecaption{ ~ \\ [-2mm] Weak and Strong $W_0^{\lambda2796}$ Comparison Samples Statistics}
\tablehead{ \colhead{Ion} & \colhead{$\bar{z}$} & \colhead{Detections} & \colhead{Upper Limits\tablenotemark{a}\tablenotemark{b}} & \colhead{Min $W_{\hbox{Ion}}$} & \colhead{Max $W_{\hbox{Ion}}$} \\   \colhead{} & \colhead{} & \colhead{} & \colhead{} & \colhead{(\AA)} & \colhead{(\AA)} }
\startdata
 ~ \\ [-2mm] \multicolumn{6}{c}{$0.3\angmath\leq W_0^{\lambda2796}\leq1.0\angmath$, $z<2$} \\ [1mm]
\hline
 ~ \\ [-2mm] HI 1215 & 0.764 & 11 & 1 & $1.09\pm0.000$ & $11.15\pm1.09$ \\
MgII 2796 & 1.069 & 170 & 0 & $0.30\pm0.04$ & $1.00\pm0.06$ \\
MgI 2852 & 1.063 & 12 & 151~(4) & $<0.05$ & $0.40\pm0.07$ \\
FeII 2374 & 1.144 & 9 & 98~(4) & $<0.07$ & $0.59\pm0.05$ \\
FeII 2586 & 1.112 & 23 & 105~(2) & $<0.03$ & $0.64\pm0.13$ \\
\enddata
\tablenotetext{a}{Upper Limits are 5$\sigma$.}
\tablenotetext{b}{Values in parentheses are upper limits derived from blended lines.}
\label{tab:sample_counts_wrz}
\end{deluxetable}

\begin{deluxetable*}{ c c c c c c }
\tablewidth{0pc}
\tablecaption{ ~ \\ [-1mm] Two Sample Tests, $W_{\hbox{Ion}} / W_0^{\lambda2796}$: Two Redshift Bins}
\tablehead{ \colhead{Ion} & \colhead{Median,} & \colhead{Median,} & \colhead{$P_{LR}$\tablenotemark{a}} & \colhead{$P_G$\tablenotemark{b}} & \colhead{$P_{PP}$\tablenotemark{c}} \\ [2pt] \colhead{} & \colhead{$0<z<2$} & \colhead{$z\geq 2$} & \colhead{(\%)} & \colhead{(\%)} & \colhead{(\%)} }
\startdata

\hi 1215 & $~~2.82$ & $~~8.86$ & $<0.1$ & $<0.1$ & $<0.1$ \\
\mgi 2852 & $~~0.15$ & $~~0.07$ & $<0.1$ & $<0.1$ & $<0.1$ \\
\feii 2374 & $~~0.17$ & $~~0.14$ & ~~~4.4 & ~~~2.4 & ~~~1.9 \\
\feii 2586 & $~~0.29$ & $~~0.20$ & ~~~1.2 & ~~~1.1 & ~~~0.8 \\
\feii 2600 & $~~0.42$ & $~~0.39$ & ~~~38.6 & ~~~51.6 & ~~~47.6 \\
\cii 1334 & $~~0.54$ & $~~0.42$ & ~~~77.5 & ~~~32.1 & ~~~34.0 \\
\civ 1548 & $~~0.67$ & $~~0.42$ & ~~~3.7 & ~~~22.0 & ~~~15.4 \\
\civ 1550 & $~~0.43$ & $~~0.27$ & ~~~0.9 & ~~~9.3 & ~~~5.7 \\
\siii 1526 & $~~0.27$ & $~~0.28$ & ~~~13.6 & ~~~51.5 & ~~~46.6 \\
\siiv 1393 & $~~0.39$ & $~~0.58$ & ~~~65.5 & ~~~34.2 & ~~~39.8 \\
\alii 1670 & $~~0.24$ & $~~0.27$ & ~~~52.8 & ~~~95.6 & ~~~75.5 \\
\aliii 1854 & $~~0.18$ & $~~0.04$ & $<0.1$ & $<0.1$ & $<0.1$
\enddata
\tablenotetext{a}{Log Rank Probability.}
\tablenotetext{b}{Gehan Probability.}
\tablenotetext{c}{Peto-Prentice Probability.}
\nonumber
\label{tab:asurv2}
\end{deluxetable*}

\begin{deluxetable*}{ c c c c c c c c c c c c c }
\tablewidth{0pc}
\tablecaption{ ~ \\ [-1mm] Two Sample Tests, $W_{\hbox{Ion}} / W_0^{\lambda2796}$: Three Redshift Bins} 
\tablehead{ \colhead{Ion} & \colhead{Median,} & \colhead{Median,} & \colhead{Median,} & \colhead{$P^{LM}_{LR}$\tablenotemark{a}\tablenotemark{b}} & \colhead{$P^{LM}_G$\tablenotemark{c}\tablenotemark{b}} & \colhead{$P^{LM}_{PP}$\tablenotemark{d}\tablenotemark{b}} & \colhead{$P^{LH}_{LR}$\tablenotemark{a}\tablenotemark{b}} & \colhead{$P^{LH}_G$\tablenotemark{c}\tablenotemark{b}} & \colhead{$P^{LH}_{PP}$\tablenotemark{d}\tablenotemark{b}} & \colhead{$P^{MH}_{LR}$\tablenotemark{a}\tablenotemark{b}} & \colhead{$P^{MH}_G$\tablenotemark{c}\tablenotemark{b}} & \colhead{$P^{MH}_{PP}$\tablenotemark{d}\tablenotemark{b}} \\ [4pt]  \colhead{} & \colhead{$0<z<1.5$} & \colhead{$1.5\leq z<3$} & \colhead{$z\geq3$} & \colhead{(\%)} & \colhead{(\%)} & \colhead{(\%)} & \colhead{(\%)} & \colhead{(\%)} & \colhead{(\%)} & \colhead{(\%)} & \colhead{(\%)} & \colhead{(\%)}}
\startdata

\hi 1215 & $~~2.87$ & ~~.......\tablenotemark{e} & $~~8.29$ & ~~....... & ~~....... & ~~....... & $<0.1$ & $<0.1$ & $<0.1$ & ~~....... & ~~....... & ~~....... \\
\mgi 2852 & $~~0.15$ & ~~0.08 & $~~0.05$ & ~~~7.3 & ~~~10.4 & ~~~8.6 & $<0.1$ & ~~~0.1 & $<0.1$ & ~~~2.1 & ~~~2.0 & ~~~1.9 \\
\feii 2374 & $~~0.21$ & ~~0.17 & $~~0.13$ & ~~~31.2 & ~~~24.0 & ~~~19.6 & ~~~3.2 & ~~~0.5 & ~~~0.5 & ~~~10.0 & ~~~7.7 & ~~~6.3 \\
\feii 2586 & $~~0.30$ & ~~0.25 & $~~0.20$ & ~~~21.6 & ~~~5.9 & ~~~8.5 & ~~~11.6 & ~~~4.8 & ~~~5.5 & ~~~41.2 & ~~~47.9 & ~~~42.5 \\
\feii 2600 & $~~0.41$ & ~~0.46 & $~~0.39$ & ~~~78.8 & ~~~91.2 & ~~~99.5 & ~~~64.5 & ~~~44.5 & ~~~51.2 & ~~~46.1 & ~~~52.4 & ~~~50.4 \\
\cii 1334 & $~~0.61$ & ~~0.44 & $~~0.42$ & ~~~2.0 & ~~~4.3 & ~~~4.3 & ~~~12.3 & ~~~10.5 & ~~~9.5 & ~~~65.9 & ~~~96.8 & ~~~95.3 \\
\civ 1548 & $~~0.72$ & ~~0.48 & $~~0.42$ & ~~~7.6 & ~~~22.2 & ~~~18.6 & ~~~8.1 & ~~~26.6 & ~~~20.6 & ~~~95.1 & ~~~75.7 & ~~~74.1 \\
\civ 1550 & $~~0.62$ & ~~0.37 & $~~0.28$ & ~~~6.2 & ~~~33.8 & ~~~22.7 & ~~~3.0 & ~~~7.3 & ~~~6.5 & ~~~46.0 & ~~~56.7 & ~~~50.5 \\
\enddata
\tablenotetext{a}{Log Rank Probability.}
\tablenotetext{b}{Superscripts represent samples compared: `L', `M', and `H' are the low, medium, and high redshift samples, respectively.}
\tablenotetext{c}{Gehan Probability.}
\tablenotetext{d}{Peto-Prentice Probability.}
\tablenotetext{e}{Too few data points.}
\nonumber
\label{tab:asurv3}
\end{deluxetable*}

\begin{deluxetable*}{ c c c c c c }
\tablewidth{0pc}
\tablecaption{ ~ \\ [-1mm] Two Sample Tests, $W_{\hbox{Ion}} / W_0^{\lambda2796}$: Weak vs Strong $W_0^{\lambda2796}$}
\tablehead{ \colhead{Ion} & \colhead{Median,} & \colhead{Median,} & \colhead{$P_{LR}\tablenotemark{a}$} & \colhead{$P_G$\tablenotemark{b}} & \colhead{$P_{PP}$\tablenotemark{c}} \\ [2pt]  \colhead{} & \colhead{$0.3\angmath\leq W_0^{\lambda2796}\leq 1.0\angmath$} & \colhead{$W_0^{\lambda2796}>1.0\angmath$} & \colhead{(\%)} & \colhead{(\%)} & \colhead{(\%)} }
\startdata

\hi 1215 & $~~6.89$ & $~~5.44$ & ~~~28.1 & ~~~44.2 & ~~~44.2 \\
\mgi 2852 & $~~0.05$ & $~~0.10$ & ~~~3.4 & ~~~2.0 & ~~~1.8 \\
\feii 2374 & $~~0.10$ & $~~0.17$ & ~~~22.6 & ~~~63.5 & ~~~33.2 \\
\feii 2586 & $~~0.18$ & $~~0.26$ & ~~~38.8 & ~~~82.5 & ~~~53.0 \\
\feii 2600 & $~~0.37$ & $~~0.47$ & ~~~4.4 & ~~~4.7 & ~~~4.1 \\
\cii 1334 & $~~0.68$ & $~~0.43$ & ~~~43.9 & ~~~12.7 & ~~~11.4 \\
\civ 1548 & $~~0.97$ & $~~0.29$ & ~~~2.7 & ~~~0.2 & ~~~0.2 \\
\civ 1550 & $~~0.70$ & $~~0.19$ & ~~~1.3 & ~~~0.1 & ~~~0.1 \\
\siii 1526 & $~~0.28$ & $~~0.30$ & ~~~95.2 & ~~~80.4 & ~~~85.2 \\
\siiv 1393 & $~~0.61$ & $~~0.19$ & ~~~0.1 & ~~~0.1 & ~~~0.1 \\
\alii 1670 & $~~0.15$ & $~~0.31$ & ~~~5.2 & ~~~10.6 & ~~~6.6 \\
\aliii 1854 & $<0.01$ & $~~0.04$ & ~~~87.3 & ~~~48.4 & ~~~94.7
\enddata
\tablenotetext{a}{Log Rank Probability.}
\tablenotetext{b}{Gehan Probability.}
\tablenotetext{c}{Peto-Prentice Probability.}
\nonumber
\label{tab:asurvwr}
\end{deluxetable*}

\begin{deluxetable*}{ c c c c c c }
\tablewidth{0pc}
\tablecaption{ ~ \\ [-1mm] Two Sample Tests, $W_{\hbox{Ion}} / W_0^{\lambda2796}$: Weak vs Strong $W_0^{\lambda2796}$ \\ ~ \\ [-2mm] Two Redshift Bins}
\tablehead{ \colhead{Ion} & \colhead{Median,} & \colhead{Median,} & \colhead{$P_{LR}$\tablenotemark{a}} & \colhead{$P_G$\tablenotemark{b}} & \colhead{$P_{PP}$\tablenotemark{c}} \\ [2pt]  \colhead{} & \colhead{$0<z<2$} & \colhead{$z\geq2$} & \colhead{(\%)} & \colhead{(\%)} & \colhead{(\%)} }
\startdata

 ~ \\ [-2mm] \multicolumn{6}{c}{$0.3\angmath\leq W_0^{\lambda2796}\leq 1.0\angmath$} \\ ~ \\ [-1mm]
\hline
 ~ \\ [-2mm] \hi 1215 & $~~2.87$ & $~~2.53$ & ~~~0.1 & $<0.1$ & $<0.1$ \\
\mgi 2852 & $~~0.10$ & $~~0.17$ & ~~~1.5 & ~~~5.6 & ~~~1.7 \\
\feii 2374 & $~~0.13$ & $~~0.18$ & ~~~31.9 & ~~~66.1 & ~~~35.3 \\
\feii 2586 & $~~0.20$ & $~~0.32$ & ~~~43.8 & ~~~41.4 & ~~~37.3 \\
\enddata
\tablenotetext{a}{Log Rank Probability.}
\tablenotetext{b}{Gehan Probability.}
\tablenotetext{c}{Peto-Prentice Probability.}
\nonumber
\label{tab:asurvwrz}
\end{deluxetable*}

\clearpage

\end{document}